\documentclass[sigconf]{acmart}

\AtBeginDocument{%
  \providecommand\BibTeX{{%
    \normalfont B\kern-0.5em{\scshape i\kern-0.25em b}\kern-0.8em\TeX}}}

\usepackage{subcaption,listings}
\usepackage{color,colortbl}
\usepackage{stfloats}
\definecolor{lightgray}{gray}{0.9}
\usepackage{ifpdf}
\usepackage{enumitem}
\usepackage{graphicx} 
\usepackage{makecell}

\copyrightyear{2026}
\acmYear{2026}
\setcopyright{cc}
\setcctype{by}
\acmConference[CHI '26]{Proceedings of the 2026 CHI Conference on Human Factors in Computing Systems}{April 13--17, 2026}{Barcelona, Spain}
\acmBooktitle{Proceedings of the 2026 CHI Conference on Human Factors in Computing Systems (CHI '26), April 13--17, 2026, Barcelona, Spain}
\acmPrice{}
\acmDOI{10.1145/3772318.3790649}
\acmISBN{979-8-4007-2278-3/2026/04}

\begin{document}
\title{Sound2Hap: Learning Audio-to-Vibrotactile Haptic Generation from Human Ratings}

\author{Yinan Li}
\orcid{0009-0000-2131-4234}
\affiliation{
    \department{School of Computing and Augmented Intelligence}
    \institution{Arizona State University}
    \city{Tempe}
    \state{AZ}
    \country{USA}
}
\email{yinanli2@asu.edu}

\author{Hasti Seifi}
\orcid{0000-0001-6437-0463}
\affiliation{
    \department{School of Computing and Augmented Intelligence}
    \institution{Arizona State University}
    \city{Tempe}
    \state{AZ}
    \country{USA}
}
\email{hasti.seifi@asu.edu}

\begin{abstract}
Environmental sounds like footsteps, keyboard typing, or dog barking carry rich information and emotional context, making them valuable for designing haptics in user applications. Existing audio-to-vibration methods, however, rely on signal-processing rules tuned for music or games and often fail to generalize across diverse sounds. To address this, we first investigated user perception of four existing audio-to-haptic algorithms, then created a data-driven model for environmental sounds. In Study 1, 34 participants rated vibrations generated by the four algorithms for 1,000 sounds, revealing no consistent algorithm preferences. Using this dataset, we trained Sound2Hap, a CNN-based autoencoder, to generate perceptually meaningful vibrations from diverse sounds with low latency. In Study 2, 15 participants rated its output higher than signal-processing baselines on both audio-vibration match and Haptic Experience Index (HXI), finding it 
more harmonious 
with diverse sounds. 
This work demonstrates a perceptually validated approach to audio-haptic translation, broadening the reach of sound-driven haptics.


\end{abstract}



\begin{teaserfigure}
    \centering
    \includegraphics[width=0.99\linewidth]{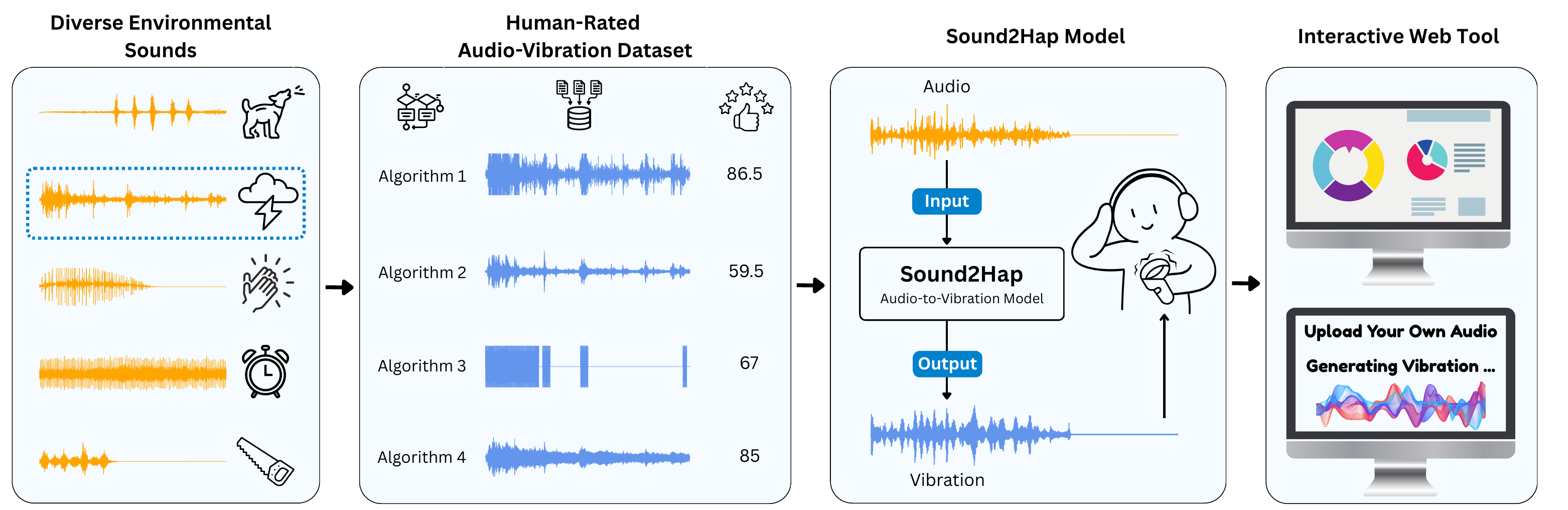}
    \caption{Overview of \textit{Sound2Hap}. Each diverse environmental sound is first converted into vibrations using four existing signal-processing algorithms. We collect human ratings for each audio-vibration pair, building a dataset of 4,000 rated samples. Using the dataset, we then train the \textit{Sound2Hap} model to generate vibrations directly from sound effects and evaluate it in a user study. In addition, we provide an online tool for visualizing the dataset and generating vibrations using both baseline signal-processing algorithms and our model, enabling designers and researchers to explore or create haptic experiences from environmental sounds.}
    \Description{A diagram showing the Sound2Hap workflow. Environmental sounds are first converted into vibration signals using four signal-processing algorithms. These audio-vibration pairs are rated by humans to build a dataset. The Sound2Hap model is then trained on this dataset to generate vibrations directly from new sound effects. The figure also shows that an online tool allows users to visualize the dataset and generate vibrations using either the baseline algorithms or the Sound2Hap model.}
    \label{fig:teaser}
\end{teaserfigure}

\maketitle
\section{Introduction}

Haptic feedback about environmental events can enhance realism, immersion, and accessibility in virtual reality (VR), multimedia, and gaming applications~\cite{krogmeier2019human, danieau2012enhancing, kim2007interactive, jiang2023collaborative}. Everyday sounds from nature, animals, and objects carry rich information and emotions, making them a valuable resource for designing haptic effects. For example, in a game or VR experience, vibrations could simulate a passing train, a barking dog, or a crackling fire to enhance immersion and support accessibility for users with visual or hearing impairments~\cite{jung2024accessible, sorgini2018haptic}. One way to achieve this is through audio-to-vibration mapping, which leverages the shared time-domain structure of both modalities. However, achieving high audio-tactile congruence is challenging due to the differences in the sensory bandwidths of the auditory and tactile systems. While hearing spans 20-20,000 Hz, touch is typically limited to frequencies below 1,000 Hz and has coarser frequency resolution~\cite{gescheider2001frequency, coren2004sensation}. These disparities make it difficult to translate fine-grained auditory features into perceptually meaningful haptics.

Prior work has introduced various signal-processing techniques to convert audio into vibrotactile patterns, typically by mapping frequency ranges~\cite{okazaki2015effect, merchel2010tactile} or perceptual features such as loudness, roughness, or pitch to vibration~\cite{lee2013real,kim2023sound}. These approaches have shown promise in targeted contexts such as enriching music listening or generating strong, discrete effects in games such as explosions or gunfire. However, their reliance on predefined mappings limits scalability and constrains haptic feedback to a narrow set of sensations. As a result, they fall short when applied to more diverse and nuanced audio content. In particular, little is known about how well such methods can translate everyday environmental sounds to vibrations, including their subtle informational and emotional cues.


To explore this, we conducted a study to gather user ratings on the perceptual match between a large, diverse set of audio-vibration pairs (Figure~\ref{fig:teaser}). We implemented four existing audio-to-vibration algorithms from the literature~\cite{sung2025hapticgen, kim2023sound, okazaki2015effect, lee2013real} and converted 1,000 sound clips from the ESC-50 environmental sound dataset~\cite{piczak2015dataset} into vibrations. These clips spanned 50 classes, including animal sounds, natural soundscapes, human non-speech sounds, and both domestic and urban environmental noises. 
In the study, 34 participants held a voice-coil vibration actuator (Haptuator Redesign) between their thumb and index finger to feel vibrations and rated 4,000 audio-vibration pairs for perceptual preference and alignment. Analysis revealed that user ratings varied widely across sound types, showing that fixed algorithmic mappings have limited generalizability. We also observed relationships between sound characteristics and algorithm performance. For example, some methods performed better for impulse or rhythmic sounds, while others were more suitable for broadband or continuous sounds, indicating that each conversion method captures different sound characteristics. These insights motivated the need for a more adaptive, data-driven approach. 


Building on these insights, we developed Sound2Hap, a convolutional neural network-based autoencoder model to synthesize vibration signals directly from audio inputs. Using our human-rated dataset from the four existing algorithms, we trained two versions of the model: one using the best-rated vibration for each sound clip (Top-Pair Sound2Hap) and another using all four vibration pairs per clip, weighted by user preferences (Preference-Weighted Sound2Hap). 
In a user study with 15 participants, we evaluated these models on new, unseen sound clips against a Baseline that applied the best signal-processing method for each sound class. 
Results showed that both model versions outperformed the Baseline, achieving significantly higher signal ratings and improved scores for Harmony, Discord, and General Score on the Haptic Experience Index (HXI) questionnaire. Between the two model variants, Top-Pair Sound2Hap received more qualitative preference from participants. These findings demonstrate the model’s ability to generalize across diverse environmental sounds while aligning with user preferences. Finally, we provide an interactive web tool that allows researchers and designers to browse our audio-vibration dataset through dynamic visualizations, examine patterns in user ratings across different sounds, and generate new haptic signals for their own audio clips.

Our contributions are as follows:
\begin{itemize}
    \item A user-rated multimodal dataset with 4,000 paired audio-vibration signals across 50 classes of environmental sounds and 8,000 user ratings. To our knowledge, this is the largest audio-vibration dataset with user ratings. 
    \item Sound2Hap, an audio-to-vibration generative model that converts diverse environmental sound effects into perceptually aligned vibrotactile feedback, trained using two alternative schemes and evaluated for generation time as well as in a user study against signal-processing baselines. 
    \item An interactive web tool that allows designers to explore the user-rated dataset and convert new sound files into vibrations using Sound2Hap and prior signal-processing algorithms. 
\end{itemize}

We make the dataset, audio-to-vibration model, and web tool open-source to support future research in this area. The model is available at GitHub\footnote{\url{https://github.com/Iris1215/Sound2Hap}}. The audio-vibration dataset can be downloaded at Hugging Face\footnote{\url{https://huggingface.co/datasets/yinanli1215/Sound2Hap}}. The web tool is available online\footnote{\url{https://sound2hap.netlify.app/}}.

\section{Related Work}  

The perceptual affinity between audio and vibrations has led to the development of various audio-to-haptic methods in both research and commercial settings. These techniques have been applied to enhance immersion in gaming~\cite{bourachot2023impact, kang2023pneumatic, yun2025real} and movies~\cite{branje2014effect, ur2014vibrotactile} by synchronizing tactile effects with sound, as well as to improve accessibility by providing vibrotactile substitutes for auditory cues~\cite{segal2024socialcueswitch, de2025tactile, cavdir2020felt, goodman2020evaluating}. We categorize these approaches into (a) signal-processing techniques and (b) learning-based neural network models.


\subsection{Signal-Processing Methods} 
Previous sound-to-tactile conversion methods rely primarily on signal-processing algorithms to convert audio signals into haptic feedback. These approaches are computationally efficient with minimal delay, making them well-suited for real-time applications such as music-related haptic experiences. Since the auditory system perceives a wide frequency range (20-20,000 Hz) while tactile sensitivity is limited ($<1$ kHz), these systems typically apply filtering and spectral transformations to compress auditory features into the vibrotactile bandwidth.

Some methods map low-level features of the audio signal directly onto vibration waveforms. For systems employing a single vibration actuator, simple spectral transformation techniques have proven effective. Frequency shifting has been used to render high-frequency musical components as haptic effects~\cite{chang2005audio, okazaki2015effect}, while dual-band linear resonant actuators have been introduced to better represent treble frequencies~\cite{hwang2013real, hwang2014improved}. Sung et al.~\cite{sung2025hapticgen} convert audio waveforms to vibrations by centering the vibration frequency at 200 Hz, near the point of highest human tactile sensitivity~\cite{gescheider2001frequency}, and introduce variation by mapping low and high intensities to a range of $\pm50$ Hz around this center frequency. When multiple actuators or actuator arrays are available, richer mappings can be achieved by distributing sound energy across different frequency bands. For example, Karam et al.~\cite{karam2009designing, karam2010emoti} demonstrate this approach using a bank of bandpass filters to map multiple frequency bands to vibrotactile stimuli across various body locations. However, such a simple bandpass filter cannot effectively capture high-frequency sound features and often fails to selectively emphasize the most informative components of complex audio clips.

Lee and Choi~\cite{lee2013real} propose a mapping from the psychoacoustic features of loudness and roughness to the perceived haptic intensity and roughness, enabling selective haptic conversion without machine learning. The mapping controls the type of sound for haptic conversion based on the sound’s perceptual quality. For example, a mapping that emphasizes only perceptually loud and rough sound can render gunshots while ignoring smoother background music, making the conversion selective. Building on this approach, Li et al.~\cite{li2021towards} introduce additional psychoacoustic features including sharpness, booming, and low-frequency energy, combined with simple threshold detection to improve conversion selectivity. However, these perceptual mapping approaches face classification accuracy challenges in determining when sounds should be converted to vibration, with Lee and Choi~\cite{lee2013real} reporting up to 20\% false positive rates and potential issues with missing low-energy target sounds or generating unnecessary tactile effects for loud, rough human voices.

Signal-processing-based methods are also widely adopted in industry and consumer products. For example, gaming headsets that emphasize low-pitch sounds convert bass frequencies into vibrations~\cite{corsair2020hs60}, and gaming controllers employ sound-driven tactile feedback to enhance immersion~\cite{playstation2020astro}. Authoring platforms such as Meta Haptics Studio~\cite{meta2023studio} and bHaptics Studio~\cite{bhaptics2021studio} allow users to convert audio files into vibrotactile signals and fine-tune them through graphical interfaces. Hardware products such as the bHaptics vest further leverage multi-band audio energy to drive a large number of vibration motors distributed across the torso, enabling full-body audio-to-haptic feedback~\cite{bhaptics2021x40}.

Most signal-processing methods have distinct advantages and limitations, and are developed primarily for gaming effects or music applications. 
To assess their efficacy for converting environmental sounds, we implement four distinct methods from the literature and construct a user-rated audio-vibration dataset.

\subsection{Learning-Based Methods}
Researchers have also explored machine learning (ML) approaches for audio-to-haptic conversion. Yoshida et al.~\cite{yoshida2017vibvid} propose VibVid, an ML-based system that learns from audio, video, and acceleration data to automatically generate vibration waveforms, demonstrated on tennis videos. Zhan et al.~\cite{zhan2023method} present a generative cross-modal framework using a Residual U-Net to synthesize tactile signals directly from audio during tool-surface interactions (e.g., sliding or tapping). These works demonstrate the feasibility of generating vibration signals with machine learning, but both are constrained to specific domains, tennis and tool-surface interactions, and therefore do not generalize to the wide variability of everyday environmental sounds that our work addresses. Furthermore, these methods rely on directly collecting paired audio and vibration signals using microphones and accelerometers~\cite{strese2016multimodal}. However, capturing high-quality vibrations is often infeasible for many environmental sounds. While the growing number of large-scale audio datasets~\cite{piczak2015dataset, audioset2017, bbc_sound_effects} offers a valuable resource for haptic design, these audio datasets lack paired acceleration data, which limits the applicability of prior learning-based approaches for these datasets.

Others use ML models to identify a set of pre-defined sound events from an audio stream, then render preset vibration effects or employ signal-processing techniques to convert the detected sounds.
For instance, Yun et al.~\cite{yun2021improving} employ deep neural networks (DNNs) to detect gunfire sounds in video games and render synchronized motion effects on a motion platform, thereby enhancing 4D gameplay experiences. In a follow-up work, they extend this approach with a random forest classifier to identify suitable sounds for haptic rendering and to selectively produce multimodal tactile stimuli including both vibrations and impacts~\cite{yun2023generating}. More recently, Yun and Choi~\cite{yun2025real} present a semantic sound-to-haptic conversion system for VR gameplay, where a Long Short-Term Memory (LSTM) model classifies in-game sounds into four key events (gunfire, hits, explosions, reloads) that trigger upper-body vibration patterns on a haptic vest. While these works highlight the potential of machine learning for sound-to-haptic conversion, most focus on classifying pre-defined sound events, which limits the generalizability of the resulting models. 

Finally, recent work has developed learning-based approaches to generate haptic signals from other modalities. Generative adversarial networks have been used to synthesize texture vibrations from images or material attributes~\cite{ban2018tactgan, cai2022gan, cai2022multi, cai2021visual, cao2023vis2hap, li2019learning, ujitoko2018vibrotactile, ujitoko2020gan}. Heravi et al.~\cite{heravi2024development} propose a model that generates texture signals in real time from user actions such as force and speed, while Faruqi et al.~\cite{faruqi2025tactstyle} employ a variational autoencoder to design physical textures for 3D-printed objects. More recently, researchers have leveraged large language models (LLMs) to generate haptic signals from free-form inputs, such as generating vibrations from text prompts~\cite{sung2025hapticgen, lim2025chathap, nakayama2024method, ren2025touched}, midair haptic patterns (e.g., rotating lines) from text~\cite{stroinski2025text}, or thermal feedback from videos~\cite{nam2024automatic}. While these approaches facilitate haptic signal design, they do not generate haptics synchronized and aligned with the audio modality. 

Our approach learns a mapping from audio waveforms to vibration signals that are perceptually aligned with human experiences, enabling vibration generation from diverse environmental sounds. 

\section{Overview of Sound2Hap Design and Evaluation Process}

\begin{figure*}[t!]
  \centering
  \includegraphics[width=0.95\textwidth]{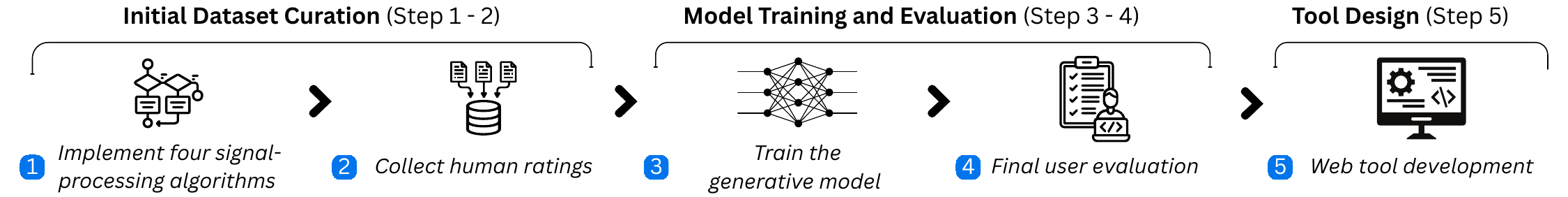}
  \caption{Our overall process for designing and evaluating Sound2Hap.}
  \Description{A flowchart with five steps showing the Sound2Hap process. Step 1: Implement four signal-processing algorithms. Step 2: Collect human ratings for the generated audio-vibration pairs. Step 3: Train the Sound2Hap model. Step 4: Conduct a final user evaluation of the model. Step 5: Develop an online web tool. The figure groups steps 1-2 as dataset curation, steps 3-4 as model training and evaluation, and step 5 as tool design.}
  \label{fig:overview}
\end{figure*}

Figure~\ref{fig:overview} illustrates our process for developing an audio-to-vibration model for environmental sounds:

\textbf{Step 1 - Implement four signal-processing algorithms (Section~\ref{sec:methods}):} We selected ESC-50~\cite{piczak2015dataset} as a diverse labeled audio dataset for environmental sounds, then implemented four existing audio-to-vibration algorithms from literature to convert 1,000 ESC-50 sound clips into vibration signals, forming the initial audio-vibration dataset.

\textbf{Step 2 - Collect human ratings (Section~\ref{sec:ratings}):} In a study with 34 participants, we collected 8,000 ratings on 4,000 audio-vibration pairs, revealing no consistent preference across algorithms and motivating a data-driven approach.

\textbf{Step 3 - Train the generative model 
(Section~\ref{sec:modeldev}):} Using the human-rated dataset, we developed Sound2Hap and trained model variants (Top-Pair and Preference-Weighted) to generate vibrations from diverse audio input, aligned with user preferences. 

\textbf{Step 4 - Final user evaluation (Section~\ref{sec:finaleva}):} A second in-person study with 15 participants showed that both Sound2Hap variants outperformed the best signal-processing baselines on perceptual ratings and HXI measures.

\textbf{Step 5 - Web tool development (Section~\ref{sec:webtool}):} We created an interactive online tool that lets designers explore the dataset and generate vibrations from new sounds using both Sound2Hap and baseline algorithms.

\section{Creating An Audio-Vibration Dataset\label{sec:methods}}
To build the dataset, we implement four audio-to-vibration conversion algorithms, each with distinct characteristics from the literature, and apply them to a dataset of environmental sound clips. 

\subsection{Audio Dataset}
As our audio source, we use the ESC-50 dataset~\cite{piczak2015dataset}, which contains 2,000 five-second environmental sound clips across five broad categories and 50 detailed semantic classes. The five sound categories include: (1) animal sounds, (2) natural soundscapes and water sounds, (3) human non-speech sounds, (4) interior domestic sounds, and (5) exterior urban environmental noises. We select this dataset due to its broad coverage of real-world environmental sounds, spanning both salient, common events (e.g., dogs barking, laughter) and more subtle, nuanced ones (e.g., brushing teeth, glass breaking). The clips emphasize foreground sound events while minimizing background noise and have clear event labeling, making the dataset well-suited for studying perceptual correspondence between audio and vibration. 
All sound clips are single-channel (mono), sampled at 44.1 kHz, and stored in 16-bit PCM format. 

To ensure consistency across all methods, we apply a normalization procedure to the sound clips. Prior to applying any audio-to-vibration conversion, all input audio signals are peak-normalized with 0 dB headroom and no additional clamping. This step ensures that waveforms are scaled to a consistent amplitude range while avoiding clipping.

\subsection{Four Signal-Processing Methods for Audio-to-Vibration Conversion\label{sec:fourtechniques}}
To create vibrations, we implement four signal-processing approaches based on prior literature and assign the following descriptive labels for ease of reference: Perception-Level Mapping~\cite{lee2013real}, Frequency Shifting~\cite{okazaki2015effect}, Pitch Matching~\cite{kim2023sound}, and HapticGen~\cite{sung2025hapticgen}. We select these four methods because they represent distinct signal-processing strategies previously used to convert audio to vibration: directly mapping perceptual features (e.g., roughness), shifting frequency components, matching sounds to a single dominant frequency (i.e., pitch), or mapping intensity and temporal dynamics. Together, they provide a diverse set of baselines for evaluating user preferences. Below, we describe each approach and note any modifications introduced in our implementations. To ensure comparability across methods, we apply a standardized normalization procedure during vibration post-processing.

\subsubsection{Perception-Level Mapping} 
This method replicates the perceptual framework developed by Lee and Choi~\cite{lee2013real}, which maps the perceptual audio features of loudness and roughness to corresponding vibration intensity and roughness. 
The algorithm operates on 4096-sample frames to extract audio loudness ($La$) based on ISO equal-loudness contours\footnote{\url{https://www.iso.org/standard/34222.html}} and roughness ($Ra$) using Vassilakis's model for spectral peak interactions~\cite{vassilakis2007sra}. These features are then mapped to vibrotactile intensity ($Iv$) and roughness ($Rv$) using the equations proposed by the authors. The final vibration synthesis employs two sinusoidal components at fixed frequencies of 175 Hz and 210 Hz, chosen through psychophysical experiments to maximize perceptual control over intensity and roughness, with amplitudes derived from $Iv$ and $Rv$. Lee and Choi proposed two sets of parameters to convert ``game'' sounds and ``music''. We implement this method using equations for the "game" category, as it is a more relevant category for environmental sounds. 

\subsubsection{Frequency Shifting}
This algorithm is based on the work of Okazaki et al.~\cite{okazaki2015effect}, who propose frequency shifting to compress the broad spectral range of audio signals, especially music, into the limited tactile frequency range (0-1,000 Hz). They sum the original audio signal with one-octave (-12 semitones) and two-octave (-24 semitones) down-shifted versions, followed by a band-pass filter centered at 250 Hz. Their experiments demonstrate high perceived harmony, comfort, and enjoyment, particularly for music containing high-frequency components such as music box sounds. Their implementation uses proprietary software (Hayaemon\footnote{\url{http://en.edolfzoku.com/hayaemon2/}}) for pitch manipulation.

We adapt this approach by developing an open-source Python pipeline with several modifications. We generate one-octave and two-octave down-shifted versions using librosa's pitch-shifting function\footnote{https://github.com/librosa/librosa}, then combine them with the original signal using waveform addition. To address the low-frequency artifacts more effectively, we first apply a 10 Hz high-pass filter to remove frequencies that cause muffled sensations. Then, we apply a 250 Hz Butterworth band-pass filter (Q = 1.0, order 4) for the final spectral shaping. 

\begin{figure*}[b!]
    \centering
    \begin{subfigure}[b]{0.6\textwidth}
        \centering
        \includegraphics[width=\linewidth]{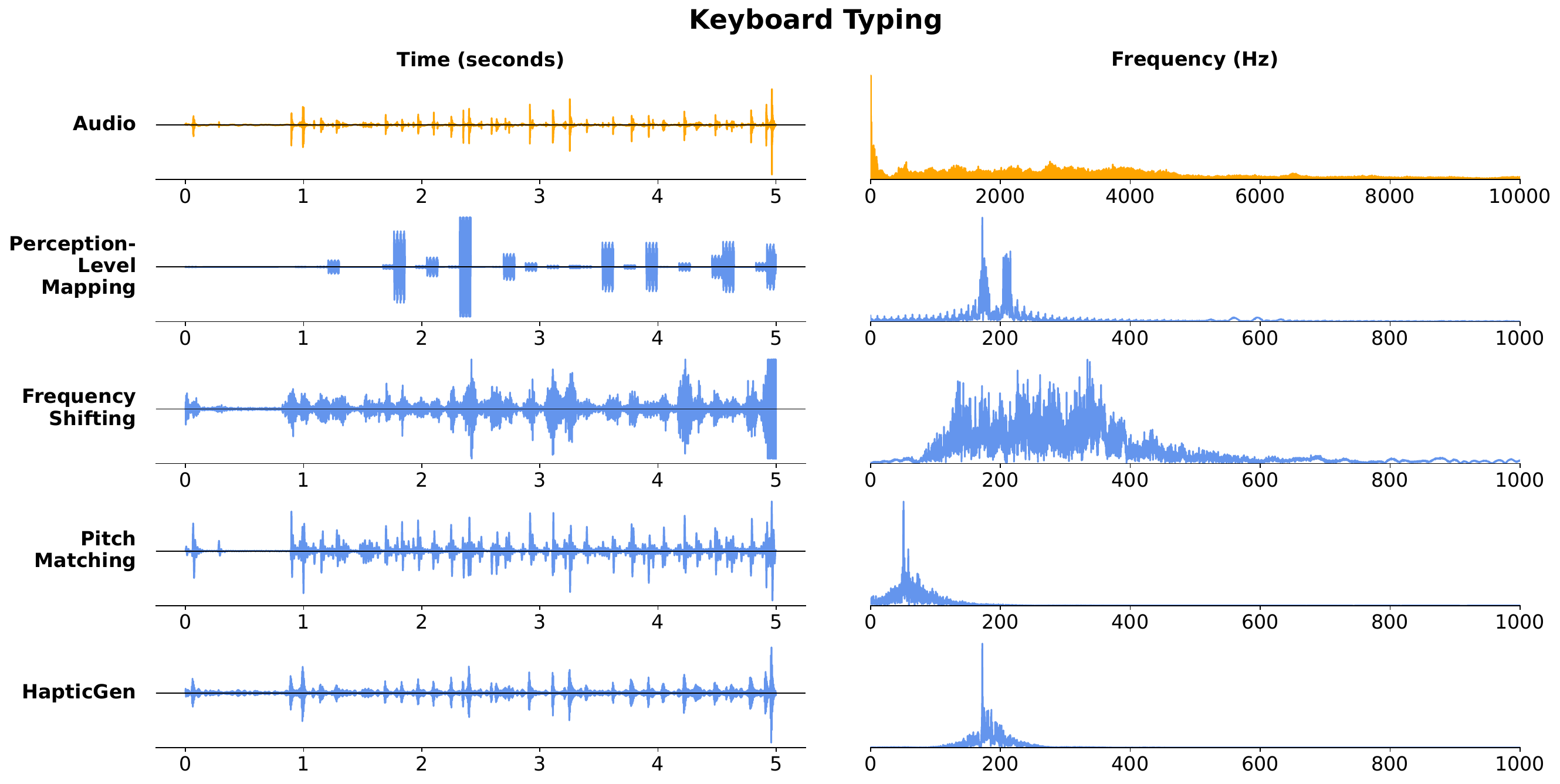}
        \caption{A \textit{keyboard typing} sound clip and vibrations in time and frequency domains.}
        \label{fig:keyboard}
    \end{subfigure}
    \hfill
    \begin{subfigure}[b]{0.6\textwidth}
        \centering
        \includegraphics[width=\linewidth]{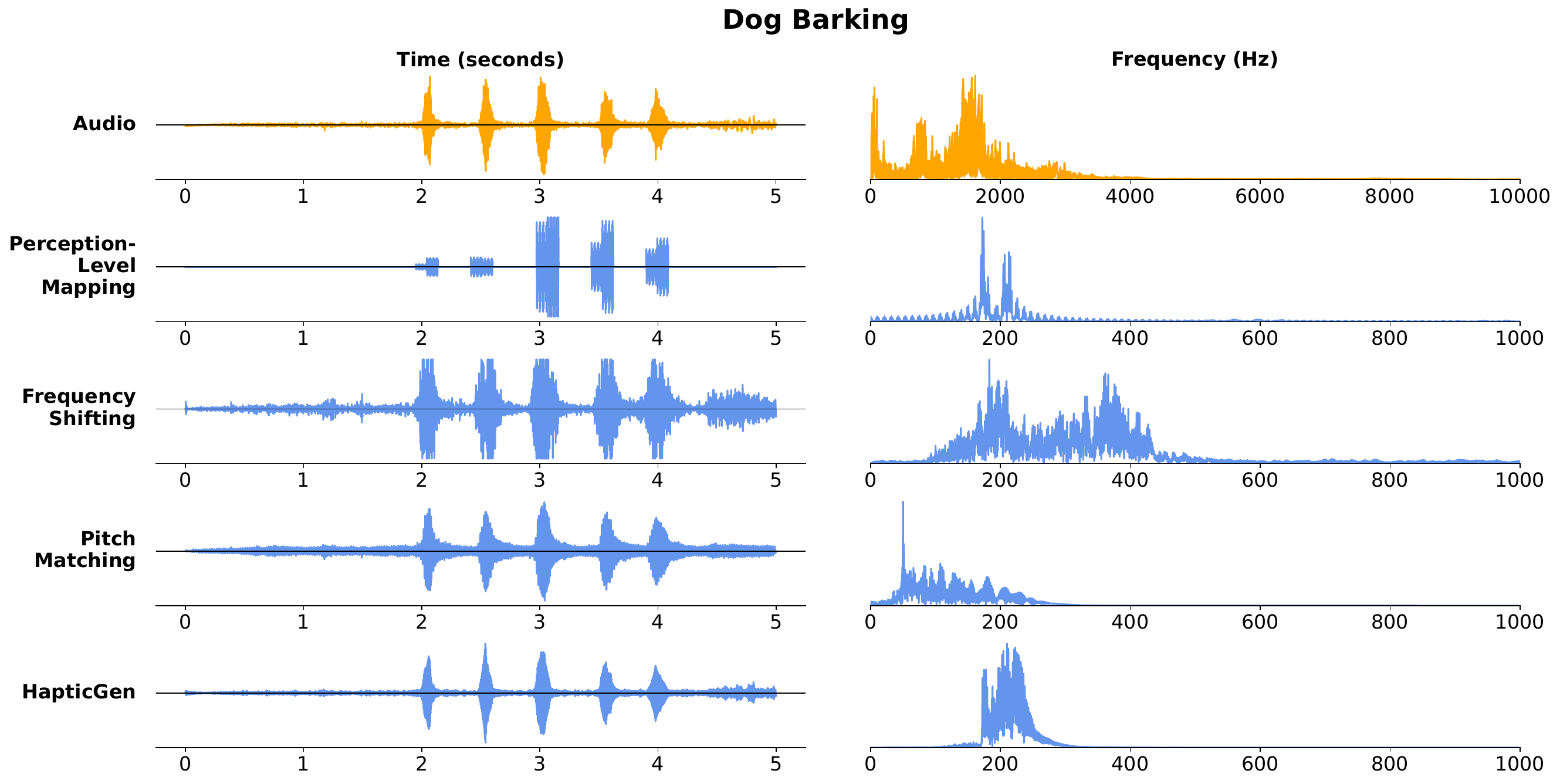}
        \caption{A \textit{dog barking} sound clip and vibrations in time and frequency domains.}
        \label{fig:dog}
    \end{subfigure}
    \caption{Examples of original audio signals (orange) and converted vibrations (blue), using four signal-processing algorithms. The audio was recorded at 44.1 kHz, 16-bit resolution, and vibrations were rendered at 8 kHz, 16-bit resolution. The y-axis of time-domain plots (left) ranges from -1 to 1. The y-axis of frequency-domain plots (right) shows magnitude, with taller spikes indicating greater energy.}
    \Description{Two example figures compare sound clips with their converted vibrations. Subfigure A shows a keyboard typing sound. Subfigure B shows a dog barking sound. For each, the audio waveform in orange is paired with four versions of vibration signals in blue, displayed both in the time domain and frequency domain. The time-domain plots show waveforms over seconds, and the frequency-domain plots show energy magnitudes across frequency ranges, with taller spikes representing stronger vibration energy.}
    \label{fig:wavsignal}
\end{figure*}

\subsubsection{Pitch Matching} 
This algorithm is based on the work of Kim et al.~\cite{kim2023sound}, who propose a crossmodal pitch-matching function to convert short sound events ($<1$ second) into vibrotactile signals. Their approach follows two steps. First, they calculate specific loudness values across 24 Bark frequency bands, using ISO 532-1, and use these features in a regression model to predict a single vibration frequency that best matches the sound. Second, the sound's loudness profile over time modulates the amplitude of a sinusoidal vibration, making the vibration mirror the sound’s dynamic intensity. The algorithm is developed using 25 short sounds of glass breaks, gunshots, swords, hitting, and explosions, but its efficacy was not tested with users. 

To extend this method to longer and more complex environmental recordings, we implement a time-varying adaptation of the original algorithm. We segment input audio into overlapping 10 ms windows with 50\% overlap, following common vibration-processing parameters~\cite{sung2025hapticgen, lim2023can}. For each segment, we then compute both specific loudness (for frequency prediction) and auditory loudness (for amplitude modulation) using ISO 532-1. We apply the regression coefficients from Kim et al.'s perceptual study to each window's Bark-band features to derive instantaneous vibration frequencies. These time-varying frequency and amplitude parameters are then interpolated across the whole audio duration and synthesized into a continuous sinusoidal waveform using phase accumulation to ensure smooth frequency transitions. This dynamic approach preserves the core crossmodal pitch matching while extending it to handle longer, frequency-varying sounds.

\subsubsection{HapticGen}
Based on the importance of temporal rhythm in vibrotactile perception, HapticGen~\cite{sung2025hapticgen} prioritizes intensity and temporal characteristics over spectral frequency details. 
The algorithm is designed to handle noisy sound clips with overlapping sound effects and complex frequency spectra, where intensity-based mapping provides a more robust haptic signal than spectral analysis. However, its perceptual effectiveness is not evaluated in a study. The vibration signal is synthesized using a sinusoidal numerically controlled oscillator with a base frequency of 200 Hz, selected for its proximity to peak human vibrotactile sensitivity~\cite{gescheider2001frequency}. Audio signals are segmented into 10 ms analysis windows, and short-term Root Mean Square (RMS) energy is computed to drive dynamic frequency modulation within a $\pm$50 Hz range around the center frequency. This approach enables the system to capture intensity variations while maintaining a frequency range optimized for tactile perception. We use the open-source code\footnote{https://github.com/HapticGen/HapticGen} shared by the authors for this method.



\subsubsection{Normalization Process for Vibration Signals}
After vibration waveform synthesis, all methods apply Root Mean Square (RMS) normalization to ensure consistent perceptual loudness across outputs. For HapticGen, Pitch Matching, and Perception-Level Mapping, the input audio is analyzed in short segments (e.g., 10-100 ms) to generate vibration signals. We estimate segment-wise RMS values and use the maximum value across segments to compute a global normalization factor for the audio. 
Frequency Shifting generates vibrations from the full audio clip in a single pass, after which we apply a global normalization factor based on the waveform’s overall RMS value. 

All final vibration signals are single-channel (mono), saved at an 8 kHz sampling rate using 16-bit PCM encoding. Figure~\ref{fig:wavsignal} presents the waveforms of two example sound clips alongside their converted vibration signals.

\section{User Study 1: Collecting User Ratings for Audio-to-Vibration Techniques\label{sec:ratings}}
We conducted an in-person user study with 34 participants to collect human preference ratings for vibration signals generated using the above four signal-processing algorithms. Participants rated 4,000 vibrations created from 1,000 sound clips in the ESC-50 dataset, resulting in 8,000 individual ratings. 

\subsection{Study Methods}
\subsubsection{Audio-Vibration Stimuli\label{kmeans}} 
We selected 20 sound clips from each of 50 audio classes in the ESC-50 dataset (half of the dataset). To ensure acoustic diversity within each class, we extracted comprehensive feature vectors from every clip, including spectral features (centroid, rolloff, and bandwidth), energy metrics (RMS energy, zero-crossing rate, and estimated tempo in beats per minute), 13 Mel-frequency cepstral coefficients (MFCCs), and a 12-dimensional chroma vector. All features were temporally averaged across each sound clip to create static descriptors. For each sound class, we applied K-means clustering (k=10) on the feature vectors to divide the acoustic space into clusters. We then sampled 20 clips per class proportionally from these clusters to capture diverse acoustic characteristics. When proportional sampling returned fewer than 20 clips, we randomly selected from the remaining clips to reach the target count. 
Each sound clip was paired with the four vibrations generated using the four signal-processing algorithms in Section~\ref{sec:fourtechniques}.

\subsubsection{Participants} 
We recruited 34 participants (24 male, 10 female; mean age = 24 years, SD = 2.4) through online advertisements at the authors’ institution. All participants had normal or corrected-to-normal vision and hearing, and no neuropathy or skin injuries on their hands. Four participants reported haptics expertise: all of them with vibrotactile technology, three with force-feedback devices, two with mid-air haptics, and one with electrotactile stimulation. 
The other 30 participants reported minimal or no prior experience in haptics. Each participant received a \$15 cash reward for their time. The study protocol was approved by the Institutional Review Board (IRB) of the authors’ institution.

\begin{figure*}[h!]
    \centering
    \begin{subfigure}[b]{0.49\linewidth}
        \centering
        \includegraphics[width=\linewidth]{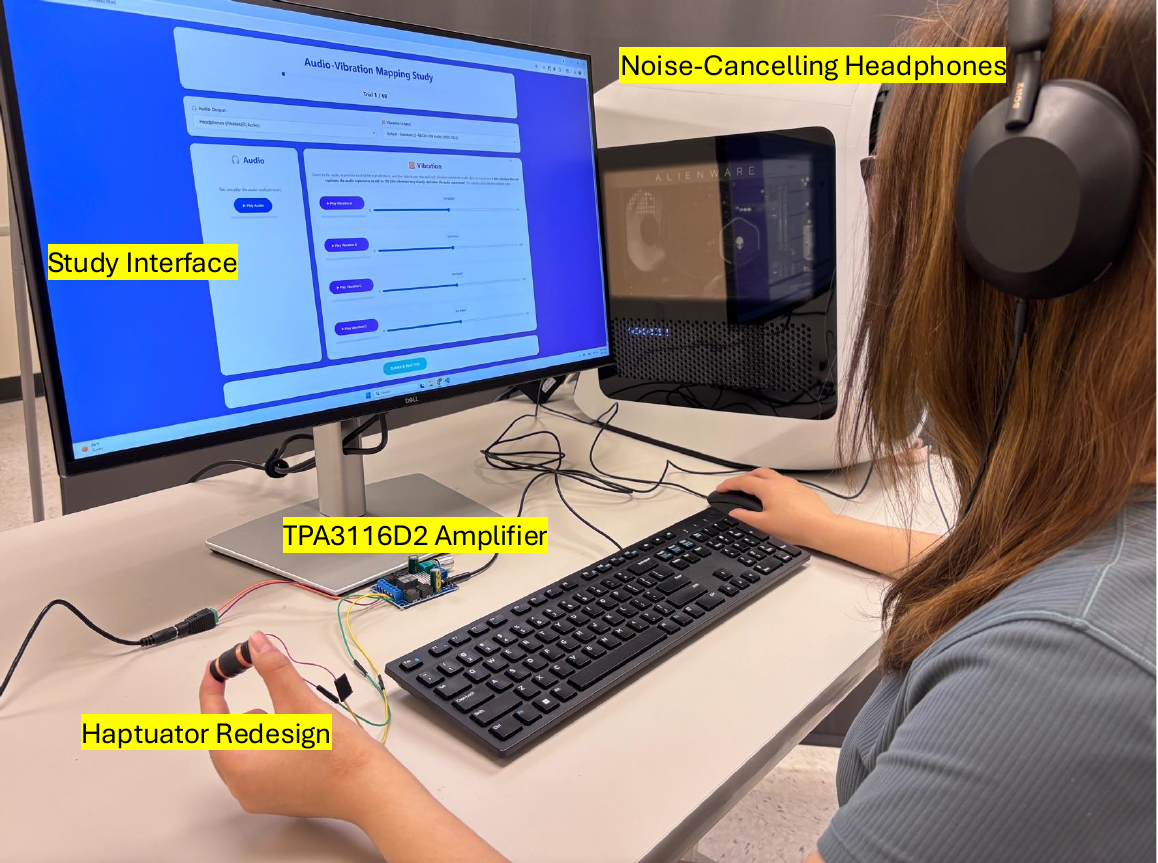}
        \caption{Study setup.}
        \label{fig:setup}
    \end{subfigure}
    \hfill
    \begin{subfigure}[b]{0.49\linewidth}
        \centering
        \includegraphics[height=6.5cm,keepaspectratio]{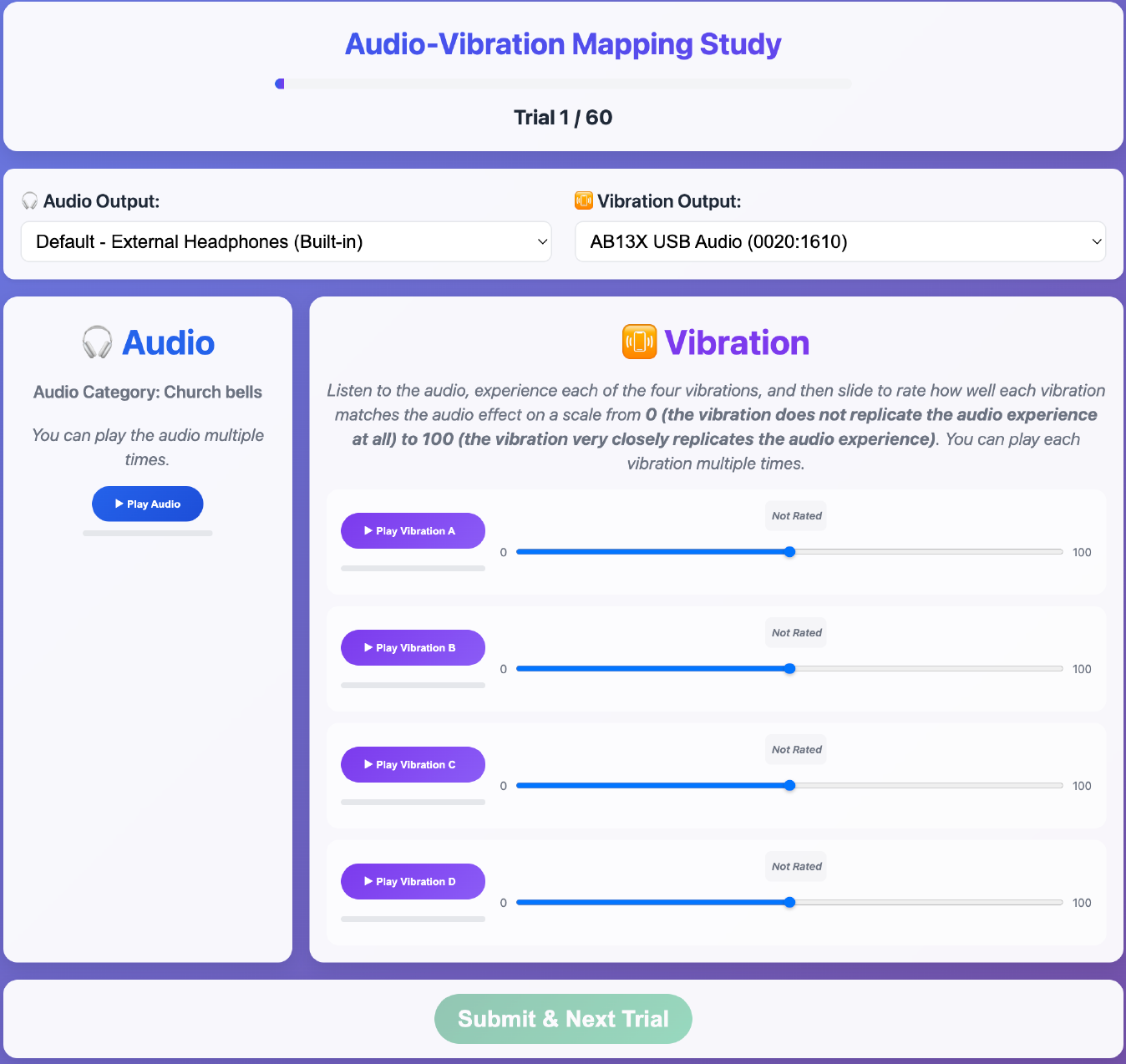}
        \caption{Study interface.}
        \label{fig:interface}
    \end{subfigure}
    \caption{Study setup and interface for the User Study 1. Participants held a voice-coil vibration actuator (Haptuator Redesign) and rated each vibration's match to the sound clips on the interface.}
    \Description{Two subfigures show the study setup and interface for User Study 1. Subfigure A depicts a participant wearing noise-cancelling headphones, holding a small voice-coil vibration actuator called the Haptuator Redesign, connected through an amplifier to a computer. The participant uses this setup to feel vibrations while listening to audio clips. Subfigure B shows the computer interface with options to play an audio clip and four corresponding vibrations. Each vibration can be rated on a scale from 0 to 100, where 0 means no match and 100 means a very close match to the audio experience.}
    \label{fig:studysetup}
\end{figure*}

\subsubsection{Procedure} 
After collecting informed consent and a background questionnaire, the experimenter introduced the study interface and instructed participants to wear noise-canceling headphones. Participants were asked to hold a voice-coil vibration actuator, namely Haptuator Redesign\footnote{\url{https://tactilelabs.com/product/tl-002-14r-haptuator-redesign/}} from TactileLabs, between the thumb and index finger of their left hand. 
The actuator was driven by a TPA3116D2 Class D stereo digital audio amplifier board connected to a 12V power supply, with its input linked directly to the PC for waveform playback. 
We calibrated audio volume, vibration intensity, and amplifier gain during a pilot study and kept them constant across all 34 participants.

Figure~\ref{fig:studysetup} shows the study setup and interface. In each trial, participants first listened to a five-second sound clip, followed by feeling four corresponding vibration signals presented in randomized order. Participants could replay both the audio and each vibration signal as many times as needed. After experiencing each vibration, they rated how well it replicated the perceptual effect of the sound on a scale from 0 (no match) to 100 (perfect match). 
Participants could only proceed to the next clip after playing the sound and four vibrations at least once and providing a rating for each vibration.

To balance clip distribution across audio classes, maintain participant engagement, and reduce fatigue, we limited each session to 60 trials. The 1,000 sound clips were divided into 17 sets: 16 sets with 60 clips each and one final set with 40 clips. Each of the first 16 sets contained one clip from every class (50 in total), plus 10 additional clips randomly selected from the remaining pool. The final set consisted of the remaining 40 clips. This design ensured that each clip received exactly two ratings, and that participants experienced a wide range of sound classes, helping maintain engagement and reduce fatigue from repeatedly rating similar clips. 
Each session lasted approximately one hour per participant.

\subsection{Results: Insights from User Ratings}
User preferences toward different algorithms varied across sound clips, classes, and categories. Below, we report preferences at the levels of five major categories, 50 sound classes, and 1,000 individual sound clips. For each individual sound clip, the rating is the average between two participants. 

\begin{figure*}[b]
    \centering
    \begin{subfigure}[b]{0.8\textwidth}
        \includegraphics[width=\columnwidth]{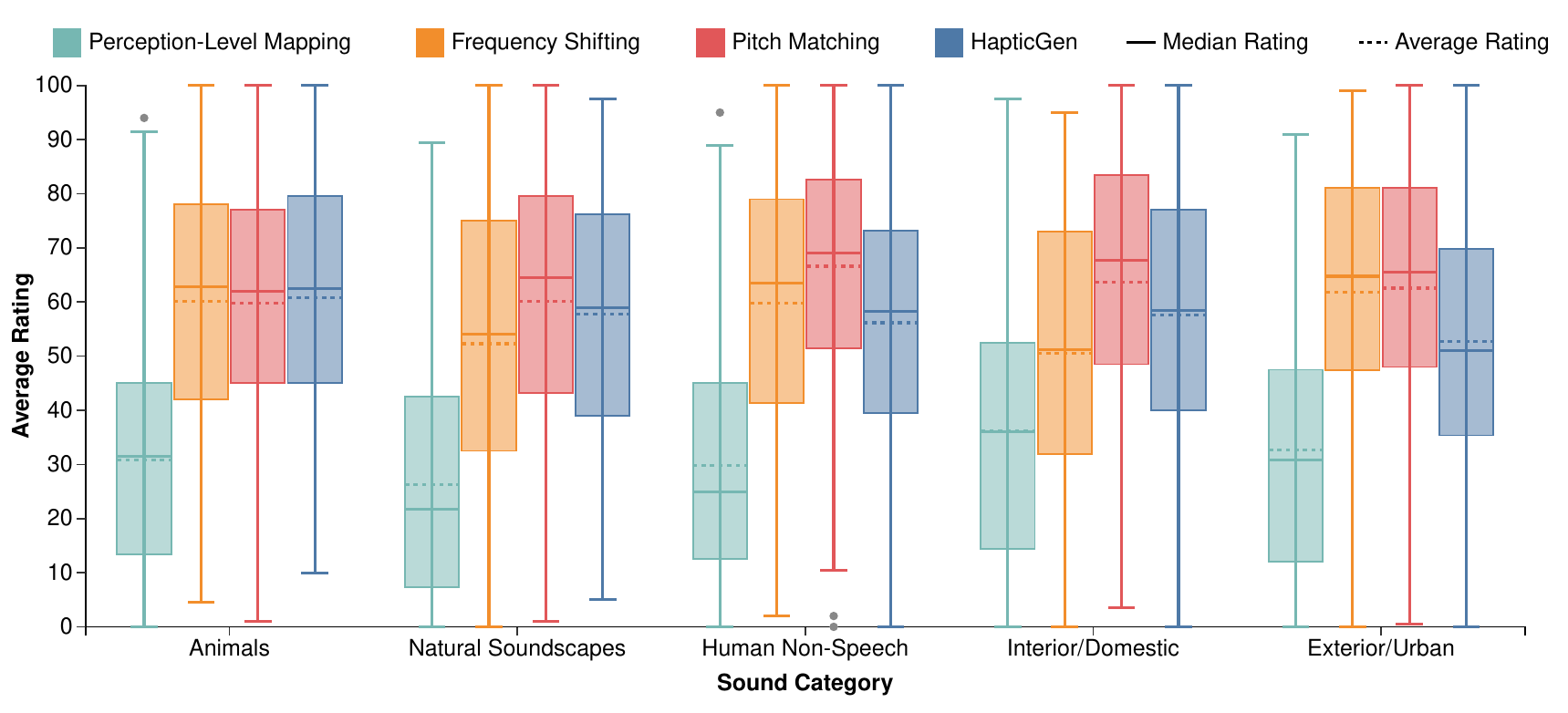}
        \caption{Boxplot of user ratings across the five sound categories for each algorithm. Each box represents 200 data points, one for each sound-vibration pair in the category.}
        \label{fig:study1box}
    \end{subfigure}
    \\[0.5cm]
    \begin{subfigure}[b]{0.8\textwidth}
        \includegraphics[width=\columnwidth]{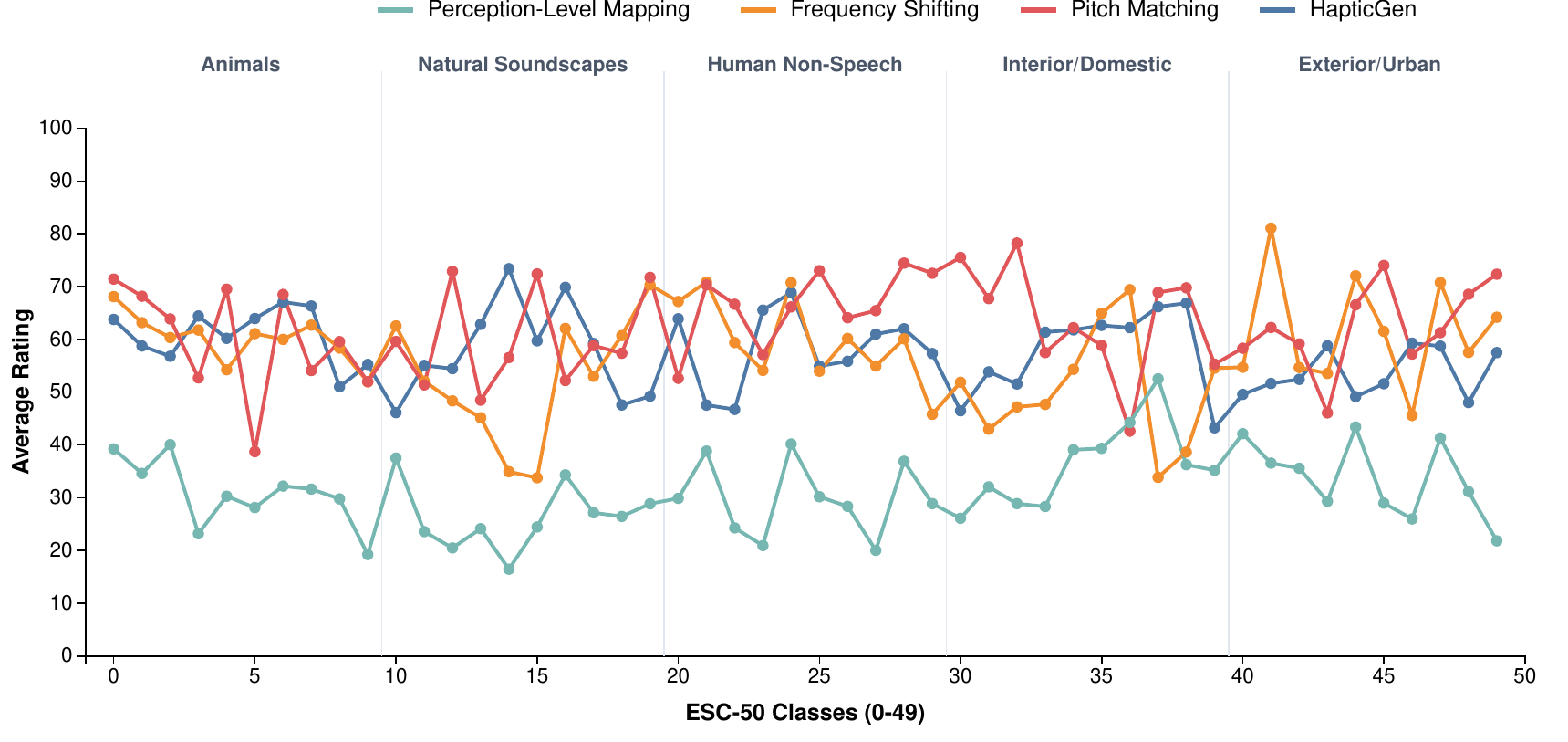}
        \caption{Average user ratings over 50 sound classes per algorithm.}
        \label{fig:study1line}
    \end{subfigure}
    \caption{Performance of the four audio-to-vibration algorithms across five overall categories and 50 sound classes.}
    \Description{Two charts compare user ratings of four audio-to-vibration algorithms. Subfigure A is a box plot showing the distribution of ratings across five sound categories—animals, natural soundscapes, human non-speech, interior or domestic sounds, and exterior or urban sounds—for each algorithm. Subfigure B is a line chart showing the average ratings for 50 individual sound classes, with separate lines representing the four algorithms. Together, the figures illustrate differences in algorithm performance across categories and detailed sound classes.}
    \label{fig:study1result}
\end{figure*}

\paragraph{Five Categories} Figure~\ref{fig:study1box} shows the distribution of clip-level ratings aggregated within each category. Each box, therefore, summarizes 200 sound-vibration pairs (10 classes $\times$ 20 clips). Ratings for each sound category and algorithm span a wide range, 
suggesting that each algorithm performs well on some sound clips but poorly on others. For \textit{Natural Soundscapes}, \textit{Human Non-Speech}, and \textit{Interior/Domestic} categories, Pitch Matching received the highest average ratings. In the \textit{Animals} category, Frequency Shifting, Pitch Matching, and HapticGen performed similarly, with no clear winner. For the \textit{Exterior/Urban} category, Frequency Shifting and Pitch Matching also showed comparable ratings.
We further averaged user ratings across all sound clips for each algorithm.
Overall, Pitch Matching achieved the highest average rating ($Mean=62.6$, $SD=22.9$ out of 100), with HapticGen ($Mean=57.0$, $SD=23.2$) and Frequency Shifting ($Mean=56.9$, $SD=24.3$) performing comparably, often close behind. 
In comparison, Perception-Level Mapping received the lowest overall ratings ($Mean=31.2$, $SD=22.9$). Overall, all methods received low or moderate average ratings for the diverse environmental sounds in our study.

\paragraph{Sound Classes} Analyzing trends across the 50 sound classes revealed more nuanced variations (Figure~\ref{fig:study1line}). Table~\ref{tab:soundclass} summarizes the winning (highest average rating across its 20 clips) algorithm for each sound class. The results suggest some links between a sound's acoustic properties and the most effective algorithm. Frequency Shifting proved more effective for continuous, noise-based, or broadband sounds with steady energy or mechanical characteristics, such as rainfall, mechanical drones (e.g., washing machine, vacuum cleaner, airplane), or engines, where conveying the overall frequency texture is more important than preserving fine temporal detail. Meanwhile, HapticGen often excelled with ambient or rhythmic soundscapes, such as sea waves, wind, crickets, and chirping birds, and sounds featuring pronounced echoing rise-and-fall envelopes 
like church bells and car horns. Finally, Pitch Matching dominated across several sound classes, especially for sequences of short, discrete events such as keyboard typing, door knocks, footsteps, and glass breaking. Both HapticGen and Pitch Matching performed well on animal sounds. 


{\renewcommand{\arraystretch}{1.5}
\begin{table*}[h!]
    \centering
    \footnotesize
    \begin{tabular}{>{\raggedright\arraybackslash}m{0.11\textwidth}  >{\raggedright\arraybackslash}m{0.12\textwidth}  >{\raggedright\arraybackslash}m{0.15\textwidth}  >{\raggedright\arraybackslash}m{0.17\textwidth}  >{\raggedright\arraybackslash}m{0.19\textwidth}  >{\raggedright\arraybackslash}m{0.12\textwidth}}
    \toprule
    \bfseries Algorithm & \bfseries Animals & \bfseries Natural\newline Soundscapes & \bfseries Human Non-Speech & \bfseries Interior/Domestic & \bfseries Exterior/Urban \\ \midrule    Perception-Level Mapping & - & - & - & - & - \\ \hline
    Frequency\newline Shifting & - & 10 rain\newline18 toilet flush & 20 crying baby\newline21 sneezing\newline24 coughing & 35 washing machine\newline36 vacuum cleaner & 41 chainsaw\newline44 engine\newline47 airplane \\ \hline
    Pitch\newline Matching & 0 dog\newline1 rooster\newline2 pig\newline4 frog\newline6 hen\newline8 sheep & 12 crackling fire\newline15 water drops\newline19 thunderstorm & 22 clapping\newline25 footsteps\newline26 laughing\newline27 brushing teeth\newline28 snoring\newline29 drinking/sipping & 30 door knock\newline31 mouse click\newline32 keyboard typing\newline34 can opening\newline37 clock alarm\newline38 clock tick\newline39 glass breaking & 40 helicopter\newline42 siren\newline45 train\newline48 fireworks\newline49 hand saw \\ \hline
    HapticGen & 3 cow\newline5 cat\newline7 insects flying\newline9 crow & 11 sea waves\newline13 crickets\newline14 chirping birds\newline16 wind\newline17 pouring water & 23 breathing & 33 door wood creaks & 43 car horn\newline46 church bells \\ \bottomrule
    \end{tabular}
    \\[0.25cm]
    \caption{Winning algorithm (rows) for each of the 50 sound classes (cells) within the five categories (columns). The number preceding each class name indicates its class number and corresponds to the x-axis in Figure~\ref{fig:study1line}. }
    \Description{A table listing which algorithm performed best for each sound class within five categories. In the Animals category, Pitch Matching was best for dog, rooster, pig, frog, hen, and sheep, while HapticGen was best for cow, cat, insects, and crow. In Natural Soundscapes, Frequency Shifting was best for rain and toilet flush, Pitch Matching for crackling fire, water drops, and thunderstorm, and HapticGen for sea waves, crickets, birds, wind, and pouring water. In Human Non-Speech, Frequency Shifting won for crying baby, sneezing, and coughing; Pitch Matching for clapping, footsteps, laughing, brushing teeth, snoring, and sipping; and HapticGen for breathing. In Interior or Domestic sounds, Frequency Shifting won for washing machine and vacuum cleaner, Pitch Matching for door knock, mouse click, keyboard typing, can opening, clock alarms, clock ticks, and glass breaking, and HapticGen for door wood creaks. In Exterior or Urban sounds, Frequency Shifting won for chainsaw, engine, and airplane; Pitch Matching for helicopter, siren, train, fireworks, and hand saw; and HapticGen for car horn and church bells. Perception-Level Mapping had no winning cases.}
    \label{tab:soundclass}
\end{table*}
}

\paragraph{Individual Sound Clips} Clip-level results revealed further insights about the algorithms. Pitch Matching was preferred as the top method for 403 clips, Frequency Shifting for 288 clips, HapticGen for 261 clips, and Perception-Level Mapping for 56 clips. For 8 sound clips, two methods tied as the highest-rated. These results show that user preferences varied depending on the specific sound characteristics. While Perception-Level Mapping had the lowest category-level and class-level average ratings, it still performed better than others for a subset of the clips. Overall, no single existing algorithm received consistently high ratings across all sounds, underscoring the need for generative models that can capture these nuances and further adapt to the diverse characteristics of audio signals. To this end, we constructed a dataset of 1,000 sound clips, 4,000 vibration signals, and 8,000 human ratings from this study to serve as training data for our model development.

\section{Sound2Hap: Audio-to-Vibration Autoencoder Model\label{sec:modeldev}} 
We developed Sound2Hap to learn an audio-to-vibration mapping directly from user perception ratings in our dataset of diverse sound effects. This section presents our dataset preparation and augmentation, model architecture and training, and evaluation of Sound2Hap's vibration generation time. 

\subsection{Dataset Preparation and Augmentation}
Our dataset contains 1000 sound samples, each paired with four corresponding vibration variants and their associated human preference ratings. We partitioned this dataset into an 80\% training set and a 20\% validation set. To ensure temporal alignment between audio and vibration signals, 
we downsampled all audio files from 44.1 kHz to 24 kHz, and upsampled the vibration signals from 8 kHz to 24 kHz using TorchAudio's FFT-based resampling\footnote{\url{https://docs.pytorch.org/audio/stable/index.html}}. 
During pre-processing, we peak-normalized input audio and normalized all target vibration waveforms to a fixed root mean square (RMS) value. This dual-normalization strategy decoupled the signal's content from its raw amplitude, producing volume-agnostic audio input and intensity-consistent vibration targets. By standardizing both domains, we preserved the statistical structure of the signals while preventing the concentration of data within a specific range. Such pre-processing was shown to improve cross-modal generation, boosting the adaptability of the model and the quality of learned signal representations~\cite{zhan2023method}. 

To improve the model's robustness and generalization, we applied common data augmentation techniques prior to training. For each sound sample, we randomly applied pitch shifting of up to $\pm2$ semitones~\cite{shakhovska2024classification, abayomi2022data} and added a small amount of Gaussian noise (up to 0.5\% of the signal's amplitude), each with a 50\% probability~\cite{moell2022speech}. This strategy enables the model to learn the core semantic features of the audio, making it robust to small variations in pitch and recording conditions and preventing overfitting. Augmentation was applied only to the training data, while the validation set was kept unchanged to track validation loss reliably, adjust the learning rate, and save the best model for evaluation. 

\subsection{Model Architecture and Training}
\begin{figure*}[b]
    \centering
    \begin{subfigure}[b]{0.95\textwidth}
        \centering
        \includegraphics[width=\linewidth]{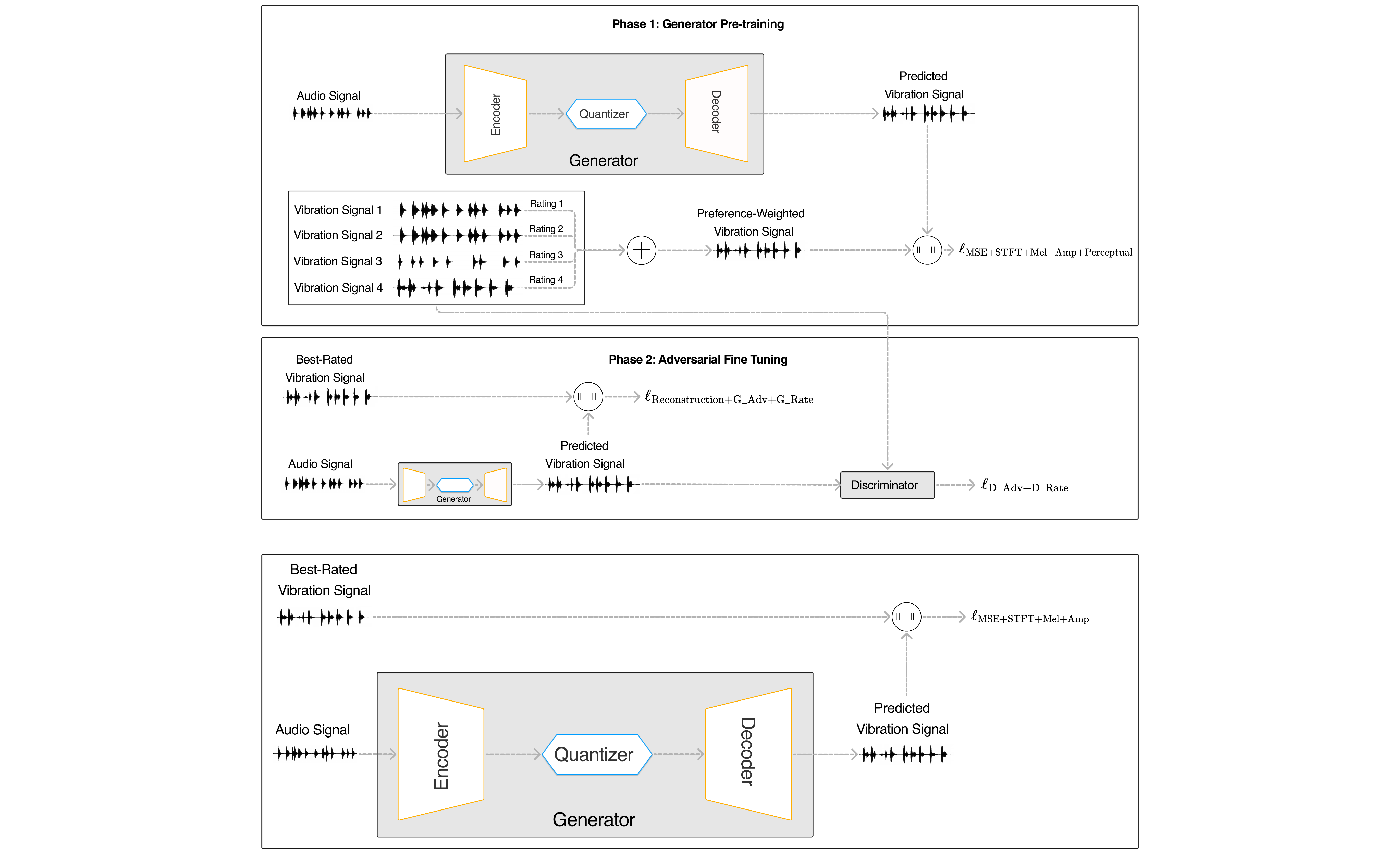}
        \caption{Top-Pair Sound2Hap}
        \label{fig:model1}
    \end{subfigure}
    \\[0.3cm]
    \begin{subfigure}[b]{0.95\textwidth}
        \centering
        \includegraphics[width=\linewidth]{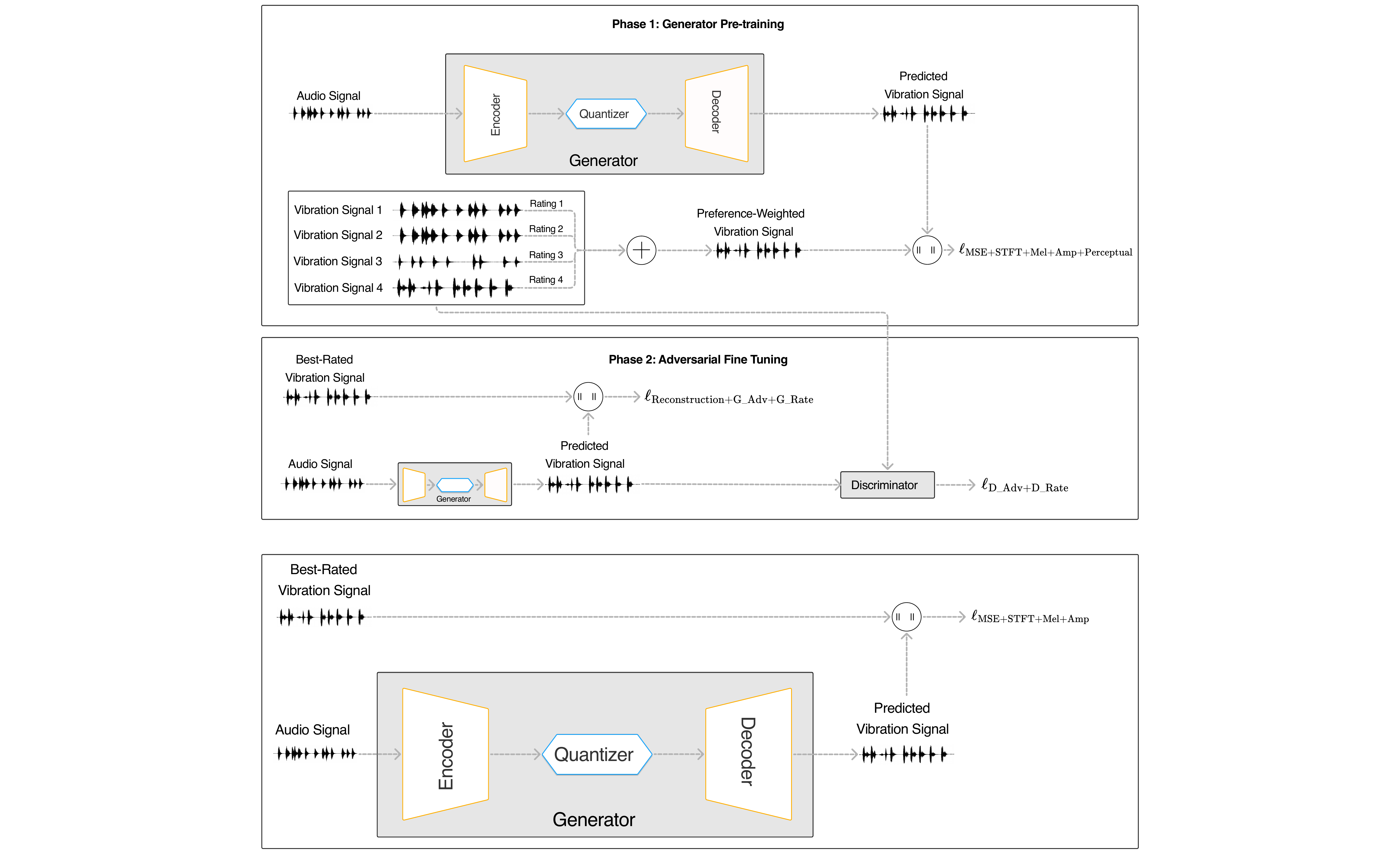}
        \caption{Preference-Weighted Sound2Hap}
        \label{fig:model2}
    \end{subfigure} 
    \caption{Sound2Hap model with two training schemes and loss functions: (a) Top-Pair Sound2Hap is trained to directly replicate the best-rated vibration signal; (b) Preference-Weighted Sound2Hap employs a two-phase training strategy: an initial pre-training phase using a blended vibration target, followed by an adversarial fine-tuning phase with a discriminator to refine the generator's output.}
    \Description{Two diagrams showing the Sound2Hap model architectures. Subfigure A, Top-Pair Sound2Hap, takes an audio signal through an encoder, quantizer, and decoder, and the generator outputs a predicted vibration signal trained to match the best-rated human vibration. Subfigure B, Preference-Weighted Sound2Hap, adds a two-phase process: first, the generator is pre-trained using a blended target created from multiple rated vibrations and their scores; then, in the second phase, adversarial fine-tuning introduces a discriminator to refine the generator’s predicted vibration output.}
    \label{fig:models}
\end{figure*}

We developed an autoencoder-based generative model for audio-to-vibration translation and trained it with two distinct strategies and loss functions (Figure~\ref{fig:models}). The architecture required an audio encoder to compress complex audio waveforms into a meaningful, low-dimensional latent representation. To achieve this, we adopted the pre-trained encoder and quantizer from the EnCodec model~\cite{defossez2022highfi}, which has demonstrated strong performance in compressing high-fidelity audio in a compact latent space.  
Our objective was to design a model that not only preserves the temporal and spectral characteristics of source audio but also incorporates subjective human preferences to generate higher-quality haptic feedback. Hence, we focused our design efforts on the decoder and the training objectives specific to the haptic domain.  

The model follows a three-stage process, with the encoder and quantizer components frozen during training to function as a fixed audio feature extractor. First, an encoder 
transforms 24 kHz audio signals into latent representations.  
We use a CNN encoder with four convolutional blocks (with residual units and downsampling) followed by two Long Short-Term Memory (LSTM) layers. 
Second, a quantizer applies residual vector quantization to discretize these latent representations into efficient codes. Third, a custom decoder translates these codes into single-channel vibration signals at 24 kHz. We use the SEANet architecture for the decoder, which has a series of residual upsampling blocks and two LSTM layers to capture the long-range temporal dependencies in vibrotactile signals. 

In our implementation, we replaced the standard ReLU activations with LeakyReLU to prevent vanishing gradients and improve training stability. Finally, we removed the typical \texttt{tanh} output activation to allow the model to learn natural vibration intensities directly rather than constraining the output amplitudes to a normalized range. 
We trained two variants of Sound2Hap under distinct loss functions and training strategies, as shown in Figure~\ref{fig:models}. 

\subsubsection{Top-Pair Sound2Hap}
The first variant, the \textit{Top-Pair Sound2Hap}, is trained using a direct, best-target strategy (Figure~\ref{fig:model1}). Its objective is to generate an output that precisely matches the single vibration that received the highest human preference rating for a given sound clip.

The training is guided by a composite loss function that compares the generated vibration against the best-rated vibration using four key metrics: (1) Mean squared error (MSE) loss supports basic time-domain waveform reconstruction; (2) Multi-resolution STFT loss compares spectrograms at different resolutions (FFT sizes of 1024, 512, 256); (3) Mel-spectrogram L1 loss emphasizes perceptually relevant frequency bands; (4) Amplitude loss adjusts the overall intensity of the prediction with the target vibration. See Supplementary Material for the loss function.

\subsubsection{Preference-Weighted Sound2Hap}
The second variant, the \textit{Preference-Weighted Sound2Hap}, learns from the full spectrum of user ratings rather than a single top choice (Figure~\ref{fig:model2}). This is achieved through a two-phase training strategy that incorporates a Generative Adversarial Network (GAN). 

\paragraph{Phase 1: Generator Pre-training} The generator's decoder is trained using a blended reconstruction loss. For each sound sample, we create a blended target vibration by computing a weighted average of all four reference vibrations, where the weights are the normalized human ratings of each vibration. This approach allows us to generate composite signals that no individual algorithm can produce, thereby enriching the dataset and mitigating limitations in sample size. This is similar to established techniques in audio processing, such as linear time-domain audio mixing used in source-separation training~\cite{hershey2016deep, cosentino2020librimix} and mixup methods that combine waveforms for regularization~\cite{tokozume2017learning, zhang2017mixup}. 
The generator minimizes a loss function between its output and this blended target, combining the MSE, STFT, Mel, and Amp losses (same as Top-Pair Sound2Hap's losses) with an additional perceptual loss term, which further improves feature-level similarity.  In this phase, the blended signals encourage the generator to learn a broader range of vibration patterns based on varied user preferences before being fine-tuned for task-specific refinement in the next phase. 

\paragraph{Phase 2: Adversarial Fine-tuning} In the second phase, we introduce a convolutional discriminator and refine the generator in an adversarial setup. The generator's training objective now combines more terms: (1) Reconstruction loss, which contains the same components as the Phase 1 objective; (2) Adversarial loss that pushes the generator to create outputs that the discriminator classifies as real; (3) Rating loss that rewards the generator for producing vibrations that the discriminator predicts will receive a high user rating. The discriminator is trained on each clip's four ``real'' vibration signals from our dataset alongside the generator's predicted signal, to classify real versus generated vibrations while simultaneously predicting user ratings for real vibrations. See Supplementary Material for the loss function and details of the convolutional discriminator.

\subsection{Sound2Hap's Vibration Generation Time}


We evaluate the model’s generation time on sound clips ranging from 1 to 20 seconds in duration. All benchmarks were conducted on 50 randomly selected 5-second, single-channel sound clips (44.1 kHz, 16-bit PCM). Each clip was segmented into five 1-second clips (250 clips total) and into two 2-second clips (ignoring the last second, 100 clips total). To simulate longer inputs, each clip was repeated to create 10-second ($\times 2$) and 20-second ($\times 4$) versions, providing an upper bound for the typical duration of vibration events in user applications. Benchmarks were run on a workstation with an Intel i7-14700HX CPU (16 GB RAM) and an NVIDIA RTX 4060 Laptop GPU (8 GB VRAM), using CUDA acceleration for model execution. The inference times reported below represent averages over all test clips: 250 clips of 1 s, 100 clips of 2 s, and 50 clips each at 5 s, 10 s, and 20 s.

Both Sound2Hap variants maintained vibration generation latencies below 1 second. 
For the Top-Pair Sound2Hap, inference required 0.44 s for 1 s and 2 s sound clips, 0.48 s for 5 s, 0.52 s for 10 s, and 0.62 s for 20 s. The Preference-Weighted variant showed similar performance: 0.44 s (1 s and 2 s), 0.52 s (5 s), 0.54 s (10 s), and 0.63 s (20 s). These results suggest that Sound2Hap maintains a stable, near-constant overhead for short clips, increasing only modestly with longer durations. In practice, vibration durations are typically limited to a few seconds to prevent overwhelming or numbing the user’s sense of touch.



\section{User Study 2: Evaluating the Sound2Hap Generative Model\label{sec:finaleva}}
We conducted an in-person user study with 15 participants to compare vibrations generated by the two Sound2Hap variants against the best-performing algorithm for each sound class from User Study 1 as a baseline. 
We selected two distinct sound sources, new sound clips from the ESC-50 dataset~\cite{piczak2015dataset} and the BBC Sound Effects library~\cite{bbc_sound_effects}, to test generalizability across different audio datasets.

\subsection{Baseline}
As a baseline, we used the best-performing signal-processing algorithm for each of the 50 sound classes. This setup mimics a ``perfect classifier'' that always detects the correct sound class and selects its best-rated method (e.g., Pitch Matching for Dog and HapticGen for Cat sounds). This baseline is motivated by prior work~\cite{yun2023generating, yun2025real}, which used a classification approach to identify relevant sounds in an audio track and play corresponding pre-designed vibrations.
In practice, we simply assign the best-performing algorithm per class in Study 1 (Table~\ref{tab:soundclass}) whenever a sound from that class appears. This approach does not involve training or performing actual classification; instead, it assumes perfect knowledge of class membership and access to user preference data per class. By always applying the empirically best-performing method for each class, this baseline represents a practical upper bound on the performance of existing signal-processing methods and serves as a strong baseline for our generative models.

\subsection{Study Methods}
\subsubsection{Audio-Vibration Stimuli}
We selected a total of 600 sound clips for evaluation from the ESC-50 dataset and the BBC Sound Effects library to convert to vibrations. 
Because the Baseline method depends on the best-performing algorithm within each class, we selected clips spanning the same 50 sound classes from both datasets. These datasets contain no shared sound clips due to their distinct sources\footnote{\url{https://freesound.org/}} and licensing terms. 

\paragraph{ESC-50 Dataset} We selected 306 clips from the remaining 1,000 clips in this dataset.  
We randomly selected six clips from each of the 50 classes. As an exception, we included 12 clips for the ``mouse clicking'' class since we did not find a matching class in the BBC dataset. 

\paragraph{BBC Sound Effects Library} We selected 294 clips from this library. For each ESC-50 class, we searched the BBC database using the exact class name as a keyword, sorting results by relevance. When exact class names did not yield relevant results, we refined our keywords with related terms. For example, using ``tap dripping'' instead of ``water drops'', or ``circular saw'' instead of ``chain saw'', to locate functionally equivalent sounds. For each class, we included the first six clips that (1) contained a clear segment of the target sound, and (2) were free of overlapping sound effects. 
All selected BBC clips were manually trimmed to the relevant sound segment. To have a uniform clip duration, segments shorter than 5s were padded, and those longer than 5s were trimmed. 

\begin{figure*}[t]
    \centering
    \begin{subfigure}[t]{0.75\textwidth}
        \centering
        \includegraphics[width=\linewidth]{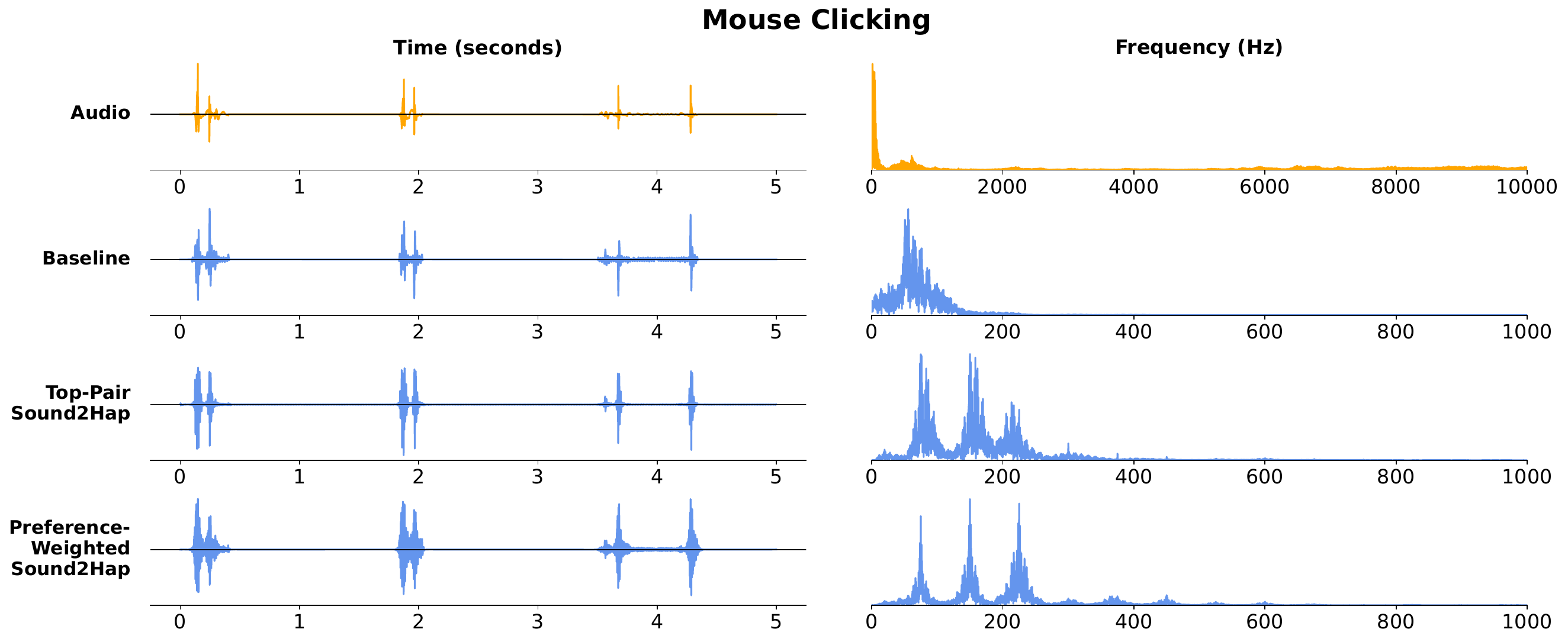}
        \caption{A \textit{mouse click} sound clip from the ESC-50 dataset and three generated vibrations in time and frequency domains. The Baseline vibration was generated using Pitch Matching.}
        \label{fig:breathing}
    \end{subfigure}
    \\[0.2cm]
    \begin{subfigure}[t]{0.75\textwidth}
        \centering
        \includegraphics[width=\linewidth]{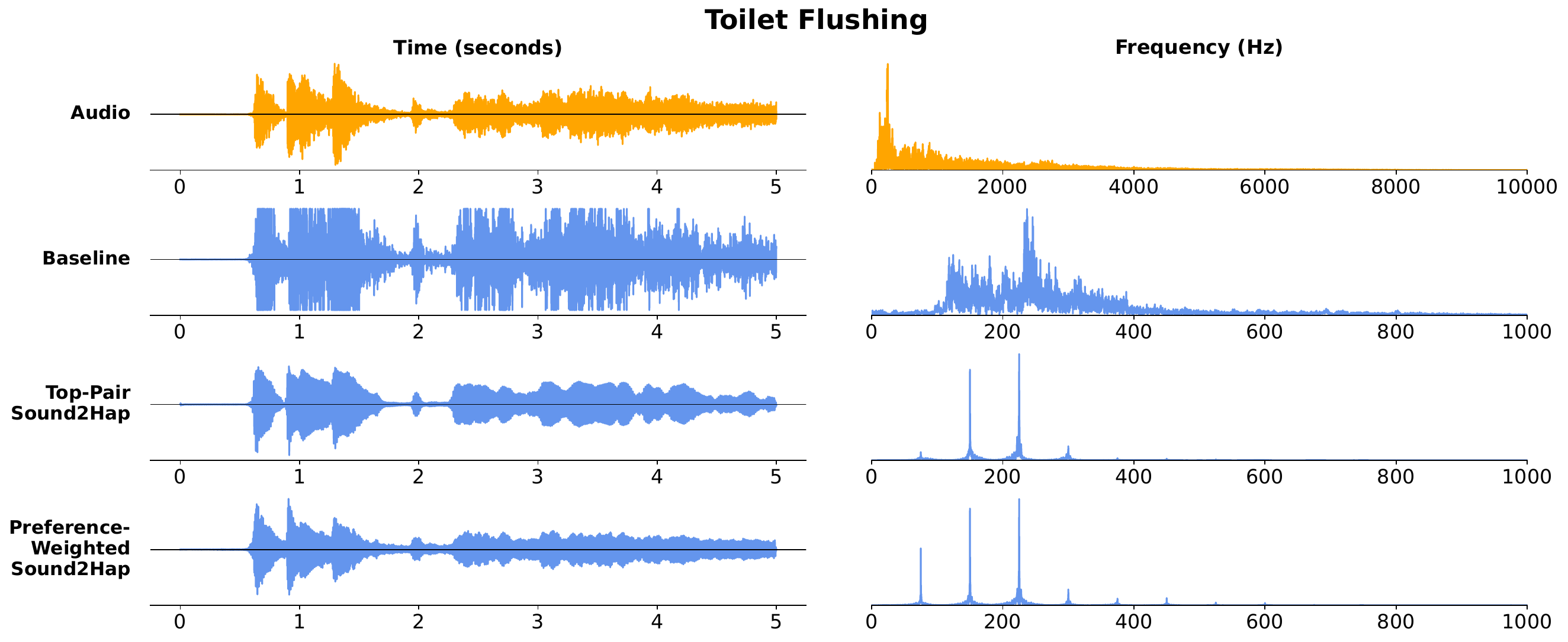}
        \caption{A \textit{toilet flush} sound clip from the BBC library and three generated vibrations in time and frequency domains. The Baseline vibration was generated using Frequency Shifting.}
        \label{fig:toilet}
    \end{subfigure}
    \caption{Two example audio signals and corresponding vibrations by Baseline, Top-Pair Sound2Hap, and Preference-Weighted Sound2Hap. For the \textit{mouse click} sound, the participant gave ratings of 69 (Baseline), 100 (Top-Pair Sound2Hap), and 79 (Preference-Weighted Sound2Hap) out of 100, and for the \textit{toilet flush} sound, the ratings were 0, 100, and 88 for the three methods, respectively.} 
    \Description{Two sets of examples comparing audio clips with generated vibrations. Subfigure A shows a mouse click sound from the ESC-50 dataset, with vibrations produced by the Baseline method using Pitch Matching, the Top-Pair Sound2Hap model, and the Preference-Weighted Sound2Hap model. The participant gave ratings of 69 for the Baseline, 100 for Top-Pair Sound2Hap, and 79 for Preference-Weighted Sound2Hap. Subfigure B shows a toilet flush sound from the BBC library, with vibrations produced by the Baseline method using Frequency Shifting, the Top-Pair Sound2Hap model, and the Preference-Weighted Sound2Hap model. The participant gave ratings of 0 for the Baseline, 100 for Top-Pair Sound2Hap, and 88 for Preference-Weighted Sound2Hap. For both sounds, time-domain waveforms and frequency-domain spectra are shown, enabling direct comparison of how each model transforms the audio into vibration signals.}
    \label{fig:wavmodel}
\end{figure*}

All audio was then converted to mono, resampled to 44.1 kHz, and saved at 16-bit depth. We created three vibrations for each clip using the Baseline method, Top-Pair Sound2Hap, and Preference-Weighted Sound2Hap. Figure~\ref{fig:wavmodel} shows the audio and vibration signals for two example sound clips.

\subsubsection{Participants}
We recruited 15 participants (10 male, 5 female; mean age = 23.9 years, SD = 1.9) through online advertisements at the authors’ institution. Participants met the same eligibility criteria as User Study 1. Six participants reported haptics expertise: all with vibrations, two with force-feedback, three with mid-air ultrasound, and one with electrotactile. All other participants indicated minimal or no prior experience with haptic systems. 
Participants received a \$15 cash reward for their time. The study was conducted under the same IRB approval as User Study 1.

\subsubsection{Procedure}
The consent form, demographic questionnaires, and introductions to hardware and interface setup followed the same protocol as User Study 1, except that the interface now presented three vibrations. 
During the main session, each participant evaluated 40 sound clips from 40 distinct sound classes, ensuring broad coverage. For each trial, participants first listened to the sound clip, then experienced three corresponding vibration signals generated by different algorithms (Baseline and two Sound2Hap variants). The vibrations were labeled as generated by Algorithm A, B, and C, with the assignment counterbalanced across participants to reduce order effects. Participants were asked to remember their impressions of each algorithm for subsequent questionnaires and interviews. They could replay the audio and vibrations as many times as needed. After experiencing each vibration, they rated how well it matched the sound using the same 0 to 100 scale as User Study 1. They could proceed only after playing the audio, feeling the vibrations, and entering ratings. 

After all trials, participants completed the Haptic Experience Inventory (HXI) questionnaire~\cite{shi2025development} for each algorithm. The HXI is a validated 20-item questionnaire that measures haptic experience across five dimensions: Autotelics, Realism, Involvement, Harmony, and Discord. The 20 items were randomized for each participant and rated 
on a 7-point Likert scale. 
At the end, we conducted a 10-minute semi-structured interview about algorithm preferences, factors influencing them, which algorithms suited specific audio categories, and any surprising outputs. 


\subsection{Results}
We present the study results for both quantitative user ratings and qualitative interview insights. 

\subsubsection{Quantitative User Ratings}
\begin{figure}[b]
  \centering
  \includegraphics[width=0.98\columnwidth]{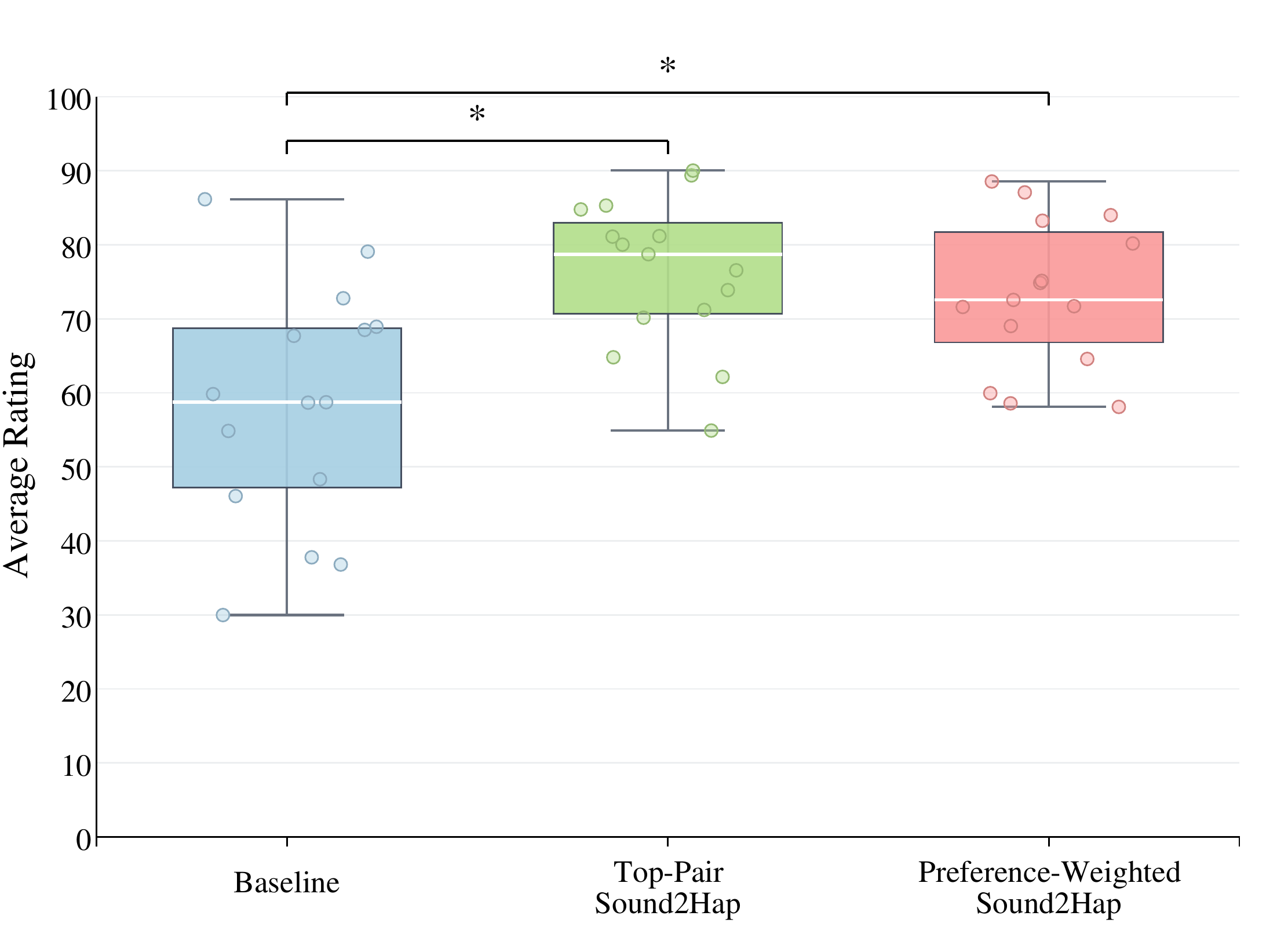}
  \caption{Ratings for audio-vibration match.}
  \Description{A box plot comparing user ratings of audio-vibration match quality across three models: the Baseline, Top-Pair Sound2Hap, and Preference-Weighted Sound2Hap. The Top-Pair and Preference-Weighted models show significantly higher ratings than the Baseline. Brackets with asterisks above the boxes indicate statistically significant differences between models.}
  \label{fig:study2rating}
\end{figure}

\paragraph{Audio-Vibration Match Ratings}  Figure~\ref{fig:study2rating} shows the distribution of ratings for the three algorithms. Across the 600 signals, average ratings were 58.28 ($SD=16.20$) for the Baseline, 76.28 ($SD=10.14$) for Top-Pair Sound2Hap, and 73.27 ($SD=10.05$) for Preference-Weighted Sound2Hap. Assumptions of normality were met, but sphericity was violated; therefore, we applied Greenhouse-Geisser correction. A one-way repeated-measures ANOVA revealed a significant effect of algorithm on signal ratings with $F(1.429,20.003)=10.466$, $p=.002$, and $\eta_p^2=.428$. Pairwise comparisons with Holm-Bonferroni correction showed that both Top-Pair Sound2Hap ($p=.011$) and Preference-Weighted Sound2Hap ($p=.013$) received significantly higher ratings than the Baseline, with no significant difference between the two Sound2Hap variants. 

The average ratings on the ESC-50 and the BBC datasets were comparable for each algorithm. The Baseline received the lowest and most varied ratings, with an average of 56.81 ($SD=33.24$) on ESC-50 and 59.80 ($SD=30.47$) on BBC. Top-Pair Sound2Hap performed best, with ratings of 76.31 ($SD=23.27$) on ESC-50 and 76.24 ($SD=24.22$) on BBC. Preference-Weighted Sound2Hap also performed well, with ratings of 72.65 ($SD=24.65$) on ESC-50 and 73.93 ($SD=24.32$) on BBC. These results suggest Sound2Hap's generalizability across sound datasets.

\begin{figure*}[b]
  \centering
  \includegraphics[width=0.95\linewidth]{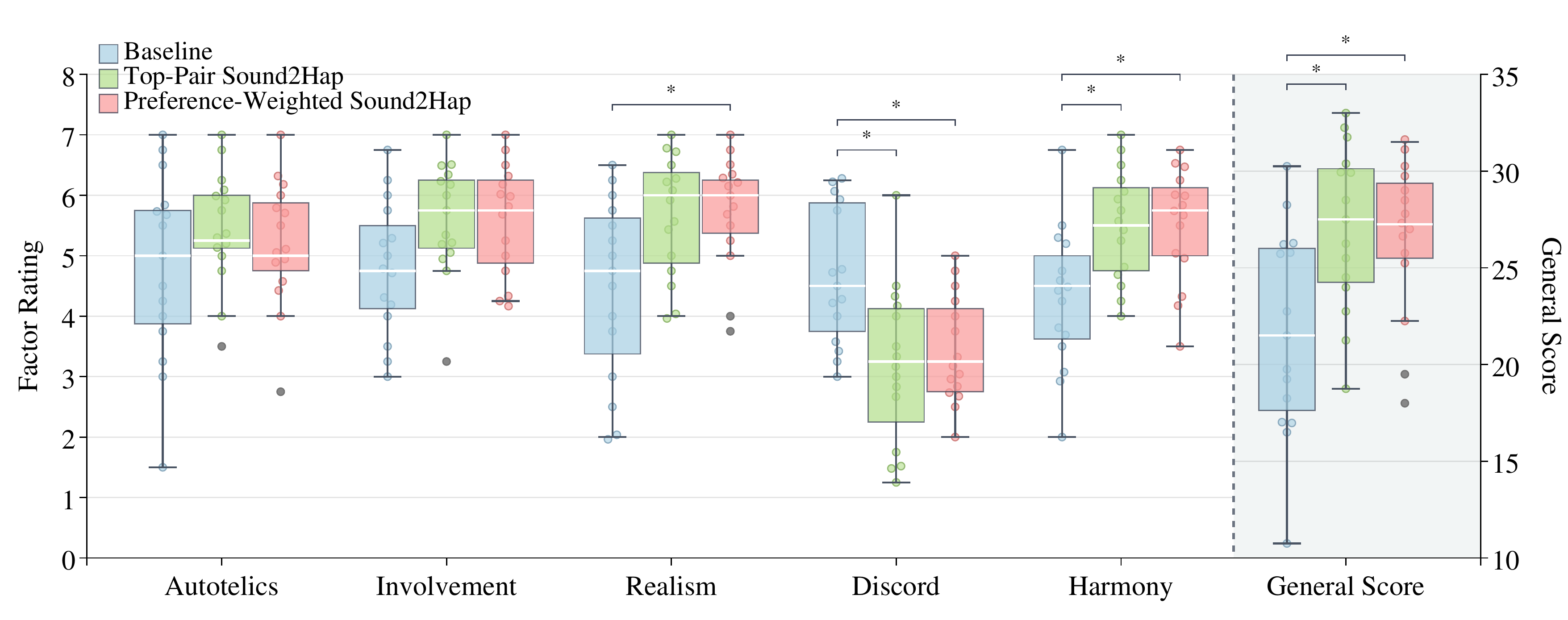}
  \caption{Algorithm ratings on Haptic Experience Index (HXI) questionnaire. Higher values reflect better performance on all factors except Discord, where lower values indicate better performance. * indicates statistical significance after Holm-Bonferroni correction. Autotelics, Involvement, Realism, Discord, and Harmony are rated on a 1 to 7 scale corresponding to the left y-axis, while the General Score, which sums these factors (reversing Discord), has a wider range corresponding to the right y-axis.}
  \Description{A box plot showing user ratings of three algorithms on the Haptic Experience Index questionnaire. Factors include Autotelics, Involvement, Realism, Discord, Harmony, and a General Score. Higher ratings indicate better performance, except for Discord, where lower values are better. Both Top-Pair and Preference-Weighted Sound2Hap outperform the Baseline on most factors. Statistically significant differences are marked for Realism, Discord, Harmony, and the General Score, as shown by brackets with asterisks above the boxes.}
  \label{fig:study2hxi}
\end{figure*}


\begin{table}[b]
    \centering
    \begin{tabular}{cccc}
         \toprule
         HXI Factor& $F$ & \textbf{$P$} & \textbf{$\eta_{p}^{2}$}\\
         \midrule
         Autotelics&1.444&.251&.093\\
         \textbf{Involvement}&\textbf{5.202}&\textbf{.027}&\textbf{.271}\\
         \textbf{Realism}&\textbf{4.999}&\textbf{.014}&\textbf{.263}\\
         \textbf{Discord}&\textbf{6.795}&\textbf{.013}&\textbf{.327}\\
         \textbf{Harmony}&\textbf{7.240}&\textbf{.003}&\textbf{.341}\\
         \textbf{General Score}&\textbf{6.007}&\textbf{.017}&\textbf{.300}\\
         \bottomrule
         \addlinespace[1.5mm]
    \end{tabular}
    \caption{Results of repeated measures ANOVA for five HXI factors and the General Score. Boldface indicates factors that were statistically significant at $p<.05$. All factors except Autothelics exhibited medium effect sizes, suggesting practical significance.} 
    \Description{A table showing results of repeated measures ANOVA for Haptic Experience Index factors. Involvement, Realism, Discord, Harmony, and the General Score all reached statistical significance with medium effect sizes. Autotelics did not show a significant effect.}
    \label{tab:anova}
\end{table}

\paragraph{HXI Ratings} Figure~\ref{fig:study2hxi} shows the distribution of HXI ratings for each algorithm. The HXI factors range from 1 (Strongly Disagree) to 7 (Strongly Agree), and the General Score is the sum of Autotelics, Involvement, Realism, Harmony, and the reverse of the score for Discord (8-Discord). The HXI ratings met the assumption of normality; however, only the Realism and Harmony factors satisfied the sphericity assumption. Therefore, we applied Greenhouse-Geisser corrections for the remaining factors and General Score. Table~\ref{tab:anova} summarizes the one-way repeated-measures ANOVA results. With the exception of Autotelics, all other factors and the General Score showed significant effects. Pairwise comparisons with Holm-Bonferroni correction revealed several significant effects. On the Realism factor, Preference-Weighted Sound2Hap ($M=5.75$, $SD=0.90$) received significantly higher ratings than Baseline ($M=4.45$, $SD=1.47$) at $p=.014$. On Discord (lower values indicate better performance), both Top-Pair Sound2Hap ($M=3.17$, $SD=1.28$) and Preference-Weighted Sound2Hap ($M=3.43$, $SD=0.75$) received significantly lower ratings than Baseline ($M=4.67$, $SD=1.21$) at the same $p=.016$. For Harmony, both Top-Pair ($M=5.52$, $SD=0.89$) and Preference-Weighted ($M= 5.52$, $SD=0.83$) were rated significantly higher than Baseline ($M=4.28$, $SD=1.30$) at $p=.018$ and $p=.005$, respectively. Finally, for the General Score, both Top-Pair ($M=27.08$, $SD=4.12$) and Preference-Weighted ($M=26.65$, $SD=3.88$) outperformed Baseline ($M=21.68$, $SD=5.19$) at $p=.025$ and $p=.016$. No other pairwise comparisons reached significance.

\subsubsection{Qualitative Interview Results}
\paragraph{Algorithm Preferences} 
When asked about their most-preferred algorithm, eight participants selected Top-Pair Sound2Hap, five preferred Preference-Weighted Sound2Hap, and two chose the Baseline. All participants reported scores between 6 and 9 on a 10-point scale regarding confidence in their selections. These qualitative preferences aligned with the user ratings. 

\paragraph{Factors Shaping Preferences} The primary factors influencing participants’ choices were the accuracy and faithfulness of the vibration patterns to the original audio. Most participants (n=8) emphasized the importance of intensity and rhythm, noting that the preferred algorithms better captured the strength and timing of the sound. Accuracy in resembling the pitch was also mentioned (n=3), while two participants highlighted overall consistency across the 40 trials. 
For example, P3 explained, \textit{“It [Top-Pair Sound2Hap] was resembling the sound more accurately than the others...the algorithm was also following the breaks and continuing the vibrations according to the intensity of the sound.”} Harmony between the audio and haptic feedback was also noted (n=4), with participants favoring vibrations that felt most synchronized with the source audio and aligned with their own perceptions and expectations. 
Top-Pair Sound2Hap was praised for its precision and ability to capture subtle sounds, rhythm, and intensity (n=8), though one user felt its vibrations were occasionally too strong. Likewise, Preference-Weighted Sound2Hap was frequently considered the best for its accuracy and consistency (n=8), but a few participants (n=4) noted it sometimes missed nuances or felt flat. Opinions on the Baseline were the most divided; while some (n=5) liked it for capturing small nuances and distinct changes, more users (n=9) found it inconsistent, confusing, and unstable compared to the other algorithms, with an intensity that was either too varied or too low.


\paragraph{Unexpected Experiences} When asked if the vibrations ever felt surprisingly different from the audio, the Baseline was the most frequent source of negative surprise (n=6). Participants described their vibrations as feeling \textit{``very off,''} \textit{``too continuous,''} or \textit{``completely different''} from what they expected. For example, P4 remarked, \textit{“Algorithm B [Baseline] would always seem really different, so I would have to go back and play the audio again [to continue rating].”} However, the Baseline also provided a few positive surprises (n=2), with one user noting it captured the intensity of a fire crackling in \textit{``an amazingly realistic way''}. In contrast, surprises from Top-Pair Sound2Hap and Preference-Weighted Sound2Hap were mostly positive (n=3), with users appreciating the nuanced sensations for sounds like a ticking clock, a swinging door, and breathing. 


\paragraph{Real-World Applicability} Looking toward real-world applications, most participants (n=13) expressed a preference for experiencing rhythmic and distinct sounds as vibrations. Popular examples included typing, clapping, thunderstorms, and various alert sounds. Some (n=5) also noted that vibrations could be engaging in specific contexts, such as gaming or watching movies, but not during music (n=2) and meditation (n=1). Conversely, nearly all participants (n=14) reported an aversion to converting continuous, loud, or unpleasant noises into vibrations. Helicopters, airplanes, chainsaws, and other jarring sounds were commonly cited as undesirable, as they were perceived as annoying or overwhelming. As one participant (P15) summarized: \textit{``Rhythmic sounds are more suitable for being converted into vibrations, whereas continuous noisy sounds feel a bit strange.”}

\subsubsection{Model-generated Vibrations Comparison with Reference Vibrations}
To evaluate why the Sound2Hap variants outperformed the Baseline, we used 200 sound clips from Study 1's validation set to analyze how closely the models reproduce vibration patterns. We compare Sound2Hap’s model-generated vibrations to two references: (a) the Baseline vibrations, produced by the winning algorithm at the sound-class level, and (b) the clip-level best vibrations, which received the highest human ratings for each clip in Study 1 (i.e., the winning algorithm for each clip). We generated vibrations for all 200 clips using both Sound2Hap variants and computed root-mean-square error (RMSE) against these two reference signals. Top-Pair Sound2Hap showed an RMSE of 0.335 from the clip-level best vibrations and 0.358 from the Baseline; Preference-Weighted Sound2Hap showed a similar pattern (0.332 vs. 0.357). These results demonstrate that both variants generate signals closer to the clip-level best vibrations than to the Baseline, supporting the conclusion that Sound2Hap captures clip-level perceptual structure beyond what class-level methods can provide.

In our subjective evaluation, we compared Sound2Hap against the class-level Baseline rather than the clip-level best vibrations, as the latter are impractical for real-world use. Identifying the best vibration for a new sound clip requires human ratings across multiple algorithms, so clip-level labels are available only in the curated Study 1 dataset, not in open-world settings (as in Study 2). Class-level knowledge, which reflects which algorithm performs best on average for a sound class, therefore represents the strongest feasible alternative for comparison.

\section{An Interactive Web Tool for Audio-Vibration Visualization and Generation\label{sec:webtool}}
\begin{figure*}[t]
    \centering
    \begin{subfigure}[t]{0.72\textwidth}
        \centering
        \includegraphics[width=\linewidth]{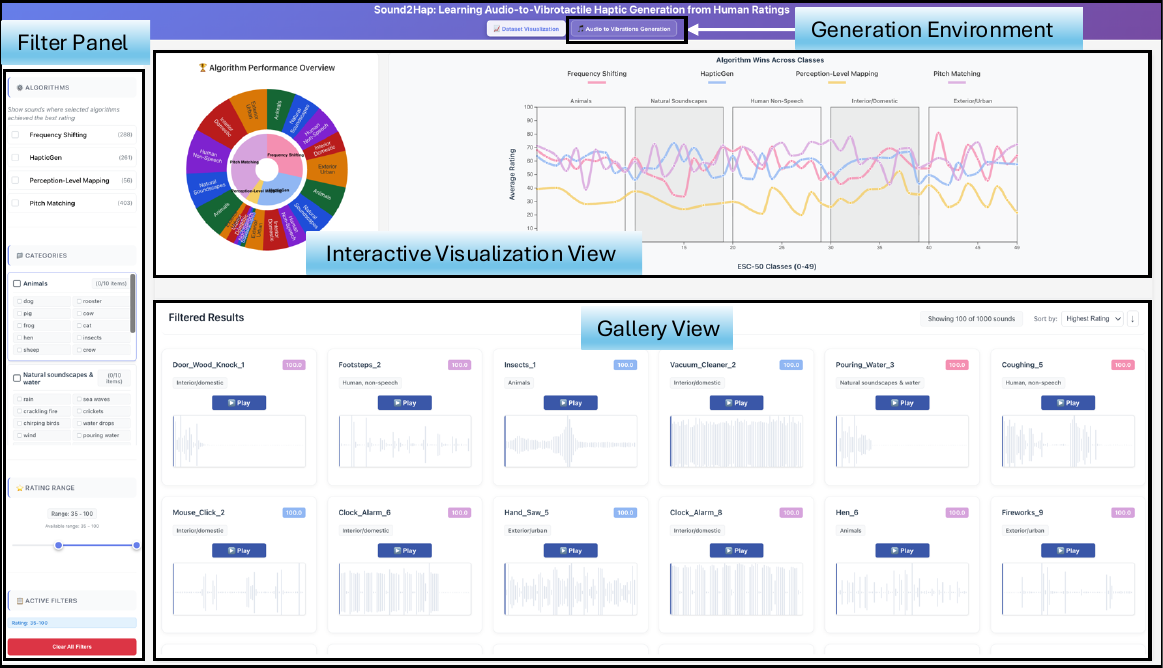}
        \caption{Interactive Visualization View (top), Gallery View (bottom), and the Filter Panel (left). The Generation Environment is accessible as a separate tab marked at the top, where users can upload audio and generate results.}
        \label{fig:tool1}
    \end{subfigure}
    \hspace{0.01cm}
    \begin{subfigure}[t]{0.26\textwidth}
        \centering
        \includegraphics[width=\linewidth,keepaspectratio]{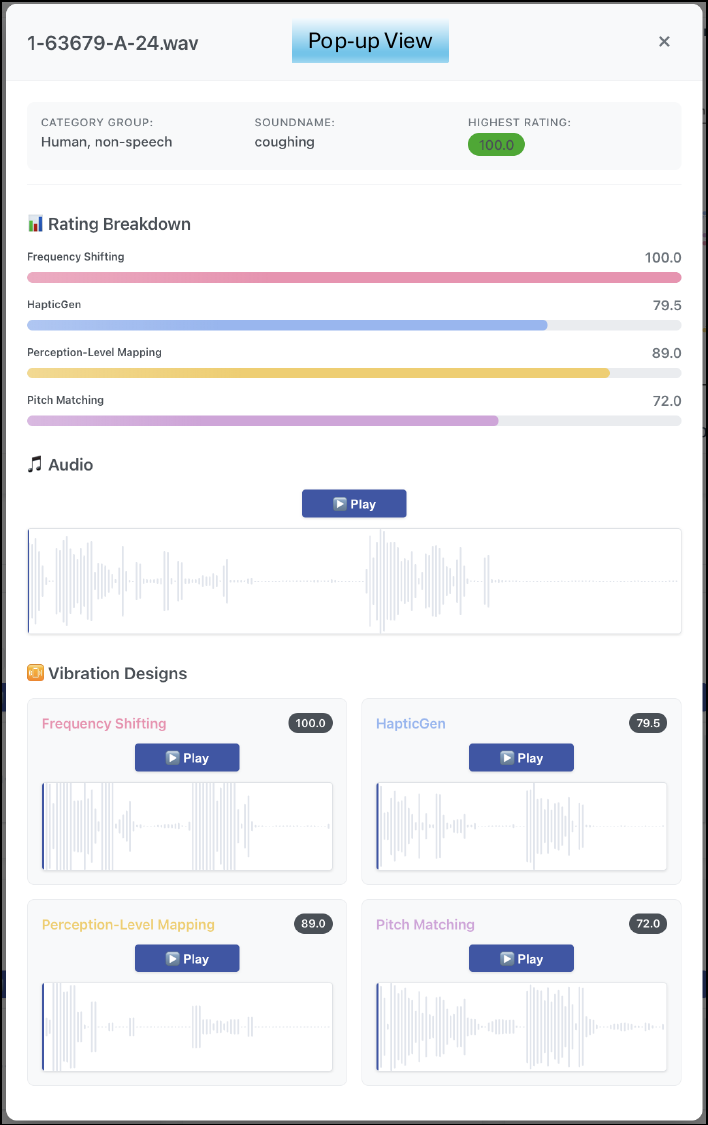}
        \caption{Pop-up View for a sound clip and corresponding vibrations.}
        \label{fig:tool2}
    \end{subfigure}
    \caption{Screenshots of the interactive web tool for audio-vibration visualization and generation.}
    \Description{Two screenshots of the Sound2Hap interactive web tool. Subfigure A shows the main interface with three sections: a filter panel on the left for selecting algorithms, categories, and rating ranges; the top visualization view with a circular chart and line chart summarizing algorithm performance; and a gallery view at the bottom showing individual audio clips with play buttons and their ratings. Subfigure B shows a pop-up view for one selected sound clip. It displays the audio waveform, a rating breakdown across four algorithms, and vibration designs with play buttons to preview each generated vibration.}
    \label{fig:tools}
\end{figure*}

We developed an online tool (Figure~\ref{fig:tools}) to facilitate access to the audio-vibration dataset and algorithms. 
The tool enables researchers and designers to browse our dataset through an interactive visualization, explore patterns of user ratings for various sounds, 
and generate new haptic signals for their audio clips. 
Specifically, the tool includes:

\begin{enumerate}[label=(\arabic*)]
    \item An \textit{Interactive Visualization View} with a sunburst chart and a line-chart, 
    allows users to explore trends across classes, categories, and conversion algorithms. 
    \item A \textit{Gallery View} at the bottom lets users browse 1,000 sound clips spanning 50 sound classes, together with their 4,000 vibration counterparts and 8,000 human ratings.
    \item A \textit{Pop-up View} shows each sound clip and corresponding vibrations with playback options, waveforms visualization, and user perception ratings.
    \item A \textit{Filter Panel} allows users to navigate the dataset efficiently by selecting subsets of sounds, categories, or conversion algorithms.
    \item A \textit{Generation Environment} lets users upload their own sound clips and generate vibrations using the four signal-processing methods and the two Sound2Hap variants introduced in this work. Generated vibrations are visualized and can be played back or downloaded for further refinement or deployment.
\end{enumerate}

\captionsetup[subfigure]{justification=centering}

\section{Discussion}
Below, we first reflect on the result, then discuss the utility of Sound2Hap and outline its limitations and future work. 

\subsection{Reflecting on Results}
\paragraph{User Perception of Audio-to-Vibration Methods} Our work is the first to directly compare user perception of multiple audio-to-vibration conversion methods across diverse sounds. Our results showed that Pitch Matching achieved the highest audio-vibration match ratings (Mean = 62 out of 100), followed by HapticGen ($M=57$) and Frequency Shifting ($M=56.9$), while Perception-Level Mapping scored lowest ($M=31.2$). 
Although each method performed well for certain clips, their average performances dropped across varied environmental sounds. In Study 2, the Baseline method, which applied the best-performing algorithm for each sound class, achieved a similar average rating ($M=58$), suggesting that class-level associations cannot fully capture the nuances of individual sounds. In contrast, Sound2Hap achieved significantly higher audio-vibration match ratings ($M=76$) and better HXI scores for Harmony, Discord, and General Score compared to the Baseline across two datasets and all sound classes. This difference can be partly explained by the training and evaluation setup. The Baseline relies on class-level averages derived from Study 1, where the winning algorithm was identified per sound class (e.g., dog barks) based on mean ratings across clips. While this provides a strong reference, it disregards the substantial clip-level variability observed within each class (e.g., different dog barks). In contrast, Sound2Hap learns directly from clip-level human ratings that encode these finer perceptual nuances, allowing the model to capture subtle relationships between acoustic features and user preference. According to Figure~\ref{fig:study1box}, Frequency Shifting, Pitch Matching, and HapticGen all received comparable mean ratings of around 60 across categories, consistent with the Baseline’s mean ratings in Study 2. However, their highest clip ratings frequently approached 100, indicating that Sound2Hap primarily learns from the top-rated vibration for each audio clip. As a result, the model generalizes more effectively to unseen sounds in Study 2, even within the same class. We further attribute this improvement to the use of an efficient low-loss audio representation in a compact latent space, as enabled by a pretrained audio encoder. 
Together, the results support the efficacy and robustness of our learning-based approach to audio-to-vibration conversion.

\paragraph{Comparison between Sound2Hap Variants} Both Sound2Hap variants performed comparably, with users showing a stronger qualitative preference for the Top-Pair model. Both variants outperformed the Baseline in user ratings and achieved similar generation times ($\sim0.5$s). Qualitative interview feedback revealed that more participants preferred the Top-Pair Sound2Hap for its precision and ability to capture rhythm and intensity, while describing the Preference-Weighted version as sometimes feeling flat or missing subtle nuances. This perceptual difference reflects the nature of their training targets. The Preference-Weighted Sound2Hap relied on a blended target that combined four vibration signals from different algorithms. 
This blended approach can introduce destructive interference, smoothing out salient details and producing less distinctive signals. By contrast, the Top-Pair Sound2Hap was trained on a single, high-quality, human-endorsed sample for each sound, providing a cleaner and perceptually coherent learning signal.

\paragraph{Spectral characteristics} As seen in Figure~\ref{fig:wavmodel}, the Sound2Hap variants learned to produce vibrations with distinct frequency spikes rather than a broad spectrum. This is likely a consequence of the training data, where two of the four source algorithms generate signals centered on specific frequencies (e.g., 175 and 210 Hz for Perception-Level Mapping, 200 ($\pm50$) Hz for HapticGen). Likewise, the Pitch Matching algorithm contributes to this characteristic by predicting a single, dominant frequency for each short audio segment, constrained to a 50 to 400 Hz range. Although this frequency varies smoothly over time, its focus on a single value at any given moment could further encourage the model to generate a narrow frequency spectrum. Thus, the model learned that preferred haptic signals often contain concentrated energy, with the highest energy frequently centered around 200 Hz, which aligns with the peak of human vibrotactile sensitivity~\cite{gescheider2001frequency}. 
From user ratings, the model learns that a few frequency spikes can adequately capture the audio nuances while feeling smooth and pleasant to touch. Moreover, this result is consistent with previous research, which demonstrates that the temporal envelope (i.e., rhythmic features) of a haptic signal dominates human perception, proving more salient than the signal's detailed spectral content~\cite{park2011perceptual, lim2024designing}.



\subsection{Utility of Sound2Hap}
\paragraph{Web Tool and Dataset} The interactive tool and dataset are designed to support exploration and reuse by researchers and designers. The visualization interface enables users to analyze the dataset by sound categories, classes, or user ratings. For example, a multimodal designer can examine, for each algorithm, the number of sound clips that received the highest ratings within a category (e.g., natural soundscapes) or class (e.g., rain, wind) or filter for highly rated audio-vibration pairs (e.g., scores > 90/100) and download them for direct use in their applications. Furthermore, the tool and open-source code enable researchers to generate large-scale vibration datasets with aligned audio and text labels from existing corpora~\cite{audioset2017, huh2025epic, chen2020vggsound}, supporting the training of sensory language models that incorporate haptics while reducing the time and cost of manual haptic data collection. By providing access to both the dataset and multiple conversion methods, the tool lowers barriers to creating high-quality, expressive haptic experiences and facilitates further model development.

\paragraph{Multimodal Applications and Research} The model’s perceptually aligned audio-based vibrations support various multimodal use cases in virtual reality, gaming, and interaction design research. For VR and video games, designers can use Sound2Hap to create expressive audio-vibration stimuli and integrate them in their virtual environments to enhance user immersion and engagement. While our results suggest that the model has a fast generation time ($\sim0.5$s) and is suitable for offline use, real-time conversion (e.g., in VR or gaming) requires stricter latency ($\sim25-75$ms)~\cite{yun2023generating, occelli2011audiotactile, adelstein2003sensitivity, zampini2005audiotactile} and would necessitate further model acceleration, such as through knowledge distillation techniques~\cite{gou2021knowledge}. Beyond applications, high-quality audio-vibration pairs provide congruent stimuli for multimodal interaction research, for example, studying how eye-free feedback modalities affect task performance in XR (e.g., gaze-based object selection)~\cite{cho2024sonohaptics}. Furthermore, prior work shows that altering environmental sounds can change self-perception~\cite{tajadura2022sensing}, such as footstep sounds modulating perceived body weight~\cite{tajadura2015light}. Our approach opens opportunities to investigate whether tactile cues for environmental sounds can produce similar psychological effects.

Moreover, the interchangeable use of audio and vibration cues or sequential playback of the two modalities used in our evaluation also reflects several real-world application scenarios. Interchangeable or substitutive use of modalities is common in alerts, notifications, accessibility cues, and assistive feedback, where either modality may be used in place of the other depending on context. For example, vibration-only notifications in silent mode or vibration replacing audio for accessibility. Sequential playback can support applications that want to notify users in stages, for example, delivering a vibration alert on a mobile phone before an audio notification to give the user an opportunity to silence the device if needed. At the same time, applications that rely on tightly synchronized audiovisual and haptic stimuli, such as immersive VR or interactive media, would require assessing concurrent playback to ensure perceptual congruence. 
See Section~\ref{sec:limit} for limitations related to playback evaluation.

\paragraph{Accessibility Applications} Sound2Hap also holds potential to support accessibility for users with visual or hearing impairments. 
For instance, perceptually congruent haptics could complement captions for environmental sounds (e.g., \textit{``[people laughing]''}, \textit{``[rain sound]''}) to enhance video accessibility for deaf and hard-of-hearing users~\cite{ahn2025enhancing}. For blind and low-vision users, such vibrations may improve video immersion~\cite{jiang2023collaborative} and support clearer audio descriptions~\cite{packer2015overview}.
As all our studies involved able-bodied participants, future work should evaluate and refine Sound2Hap with feedback from visually and hearing-impaired users to enhance its perceptual performance and assess its practical utility for accessibility applications.

\subsection{Limitations and Future Work\label{sec:limit}}
Our work has four main limitations that open up avenues for future research. First, while our dataset represents the largest and most diverse collection of everyday sound events paired with vibrations to date, real-world sounds span an even broader range. Furthermore, Sound2Hap is only tested on sound clips containing a single, salient sound source, while real-world audio often includes overlapping or blended events. Future work can assess the efficacy of Sound2Hap for a wider range of single and overlapping environmental sounds. For overlapping sounds, a promising direction is to develop a modular pipeline that integrates a pre-processing step using audio source separation~\cite{nugraha2016multichannel, liu2024separate} or saliency detection~\cite{podwinska2019acoustic} algorithms. This would allow the system to isolate a foreground event from irrelevant ones (e.g., background noise or speech) and feed the separate and cleaner audio clips to Sound2Hap, expanding its practical applicability. In addition, when audio contains multiple simultaneous sources, the model could be extended to generate multiple vibrations across a vibrotactile array (e.g., a haptic vest), with each vibration representing a distinct sound or source location. Finally, future work could adapt Sound2Hap for other sound domains relevant to interaction design, such as earcons and voice-user-interface cues. Incorporating existing datasets of interface sounds~\cite{goswami2025beepbank, cao2024bisaid} would help generalize the model to create vibrations for symbolic, short-duration sounds commonly used for UI feedback, such as haptic cues that supplement auditory feedback in noisy environments or replace it in silent modes. 

Second, our model development and evaluation focus on overall performance across users, but the quantitative ratings and qualitative preferences suggest that users may have individual differences in the algorithm they prefer. Relatedly, in our studies, the participant sample showed a gender imbalance (Study 1: 24 male / 10 female; Study 2: 10 male / 5 female), which may limit the generalizability of the findings. Since prior research suggests that gender can influence tactile simultaneity thresholds~\cite{geffen2000sex}, perceived vibration intensity and discomfort~\cite{neely2006gender}, and tactile spatial acuity~\cite{peters2009diminutive}, 
including a gender-balanced sample could reveal user differences in sensitivity to vibration signal variations or synthesis artifacts. Future work should aim to recruit more demographically balanced samples to examine whether gender or other individual factors (e.g., age, music background) impact user preference for audio-to-haptic conversion. Furthermore, future directions could explore adapting personalization approaches from other modalities~\cite{alaluf2024myvlm, schelle2015tactile, mackowski2023multimodal} to further calibrate the model's output to an individual user's perception and preferences. 

Third, our evaluation relies on sequential playback of audio and vibrations, allowing participants to focus on tactile nuances without auditory masking. This design also supports unimodal applications, such as content accessibility for users with sensory impairments. Thus, our model is optimized for audio-to-haptic translation when felt in isolation. 
Yet, many real-world applications rely on concurrent audio-haptic playback, thus our results from isolated vibration perception may not fully apply to them. 
Prior research shows that synchronizing audio and vibration may alter haptic perception; for instance, a low-frequency sound can cause a high-frequency vibration to be perceived as having a lower frequency~\cite{hassan2024audio}. Future work should therefore evaluate the holistic quality of the combined, concurrent stimulus to further optimize haptics for multimodal playback. 

Lastly, our evaluation setup focuses on finger-based perception using a voice-coil vibration actuator. This configuration enables precise and high-sensitivity evaluation of vibrotactile signals. However, tactile acuity varies across different body sites. Since the fingertip contains a high density of mechanoreceptors~\cite{johansson1979tactile}, the reported results may represent an upper bound on the perceptual gains of Sound2Hap. Also, other actuator types (e.g., eccentric rotating mass or linear resonant actuators) produce distinct vibration profiles~\cite{choi2012vibrotactile}. 
Future work should therefore examine how well Sound2Hap generalizes across actuator types, body locations, and broader usage contexts.

\section{Conclusion}
This work introduces Sound2Hap, a data-driven model for generating vibrotactile signals from diverse environmental sounds. By combining large-scale perceptual data collection with two generative model variants, we address the limitations of prior rule-based audio-to-haptics approaches. Our user study shows that Sound2Hap produces signals that align with human perceptual preferences and generalize across sound domains. We release our dataset, models, and an interactive tool to support future research and applications. We envision Sound2Hap enabling scalable haptic content creation and informing new multimodal experiences across XR, accessibility, and entertainment.

\begin{acks}
We thank Mainak Malay Saha for developing the web tool. We also thank the anonymous reviewers, our colleagues, and the study participants for their input on this project. This work was supported by research grants from the National Science Foundation (\#2339707) and VILLUM FONDEN (VIL50296).
\end{acks}

\bibliographystyle{ACM-Reference-Format}
\bibliography{references}

\end{document}